%% file: main.tex
\def\commentswitch#1{\iffalse#1\fi} 

\def\commentswitchsj#1{\iftrue#1\fi} 

\newcommand\notesj[1]{\commentswitchsj{\todo[color=orange!20, inline, size=\small]{Su: #1} }}

\documentclass[a4paper,fleqn,usenatbib]{mnras}
\usepackage[T1]{fontenc}
\usepackage{ae,aecompl}

\usepackage{graphbox,graphicx}	
\usepackage{amsmath}	
\usepackage{amssymb}	

\usepackage{dcolumn}     
\usepackage{bm}          
\usepackage{color}

\usepackage{lineno} 

\usepackage[dvipsnames]{xcolor}
\usepackage{todonotes,multirow,enumitem}
\usepackage{soul}
\usepackage[normalem]{ulem}
\usepackage{siunitx}

\usepackage{booktabs}

\usepackage{eso-pic}

\AddToShipoutPictureBG*{%
  \AtPageUpperLeft{%
    \hspace{0.75\paperwidth}%
    \raisebox{-3.5\baselineskip}{%
      \makebox[0pt][l]{\textnormal{DES-2020-0531}}
}}}%

\AddToShipoutPictureBG*{%
  \AtPageUpperLeft{%
    \hspace{0.75\paperwidth}%
    \raisebox{-4.5\baselineskip}{%
      \makebox[0pt][l]{\textnormal{FERMILAB-PUB-21-206-AE}}
}}}%

\newcommand{\ud}{\mathrm{d}}

\newcommand{\bea}{\begin{eqnarray}}
\newcommand{\eea}{\end{eqnarray}}
\newcommand{\beas}{\begin{eqnarray*}}
\newcommand{\eeas}{\end{eqnarray*}}  
\newcommand*{\swap}[2]{#2#1}

\newcommand{\hMpc}{h^{-1}{\rm\;Mpc}}

\newcommand{\degsqr}{{\rm\;deg^2}}

\newcommand{\stripeGold}{$\sim150 \degsqr$}

\newcommand{\redmagic}{redMaGiC~}
\newcommand\cosmosis{{\textsc{CosmoSIS}}}
\newcommand\cosmolike{{\textsc{cosmoLike}}}

\newcommand\metacal{{\textsc{metacalibration}}}

\newcommand\cred{\color{red}}

\def\cmt#1{{\cred#1}}

\def\citepp#1{\citeauthor{#1} \citeyear{#1}}

\title[GGL with DMASS]{Galaxy-galaxy lensing with the DES-CMASS catalogue: measurement and constraints on the galaxy-matter cross-correlation}

\author[DES Collaboration]{
\parbox{\textwidth}{
\Large
S.~Lee,$^{1}$\thanks{E-mail: sujeong.lee717@duke.edu}
M.~A.~Troxel,$^{1}$
A.~Choi,$^{2}$
J.~Elvin-Poole,$^{2,3}$
C.~Hirata,$^{2,3}$
K.~Honscheid,$^{2,3}$
E.~M.~Huff,$^{4}$
N.~MacCrann,$^{5}$
A.~J.~Ross,$^{2}$
T.~F.~Eifler,$^{6,4}$
C.~Chang,$^{7,8}$
R.~Miquel,$^{9,10}$
Y.~Omori,$^{11}$
J.~Prat,$^{7}$
G.~M.~Bernstein,$^{12}$
C.~Davis,$^{11}$
J.~DeRose,$^{13,14}$
M.~Gatti,$^{12}$
M.~M.~Rau,$^{15}$
S.~Samuroff,$^{15}$
C.~S{\'a}nchez,$^{12}$
P.~Vielzeuf,$^{10}$
J.~Zuntz,$^{16}$
M.~Aguena,$^{17,18}$
S.~Allam,$^{19}$
A.~Amon,$^{11}$
F.~Andrade-Oliveira,$^{20,18}$
E.~Bertin,$^{21,22}$
D.~Brooks,$^{23}$
D.~L.~Burke,$^{11,24}$
A.~Carnero~Rosell,$^{25,18,26}$
M.~Carrasco~Kind,$^{27,28}$
J.~Carretero,$^{10}$
F.~J.~Castander,$^{29,30}$
R.~Cawthon,$^{31}$
C.~Conselice,$^{32,33}$
M.~Costanzi,$^{34,35,36}$
L.~N.~da Costa,$^{18,37}$
M.~E.~S.~Pereira,$^{38}$
J.~De~Vicente,$^{39}$
S.~Desai,$^{40}$
H.~T.~Diehl,$^{19}$
J.~P.~Dietrich,$^{41}$
P.~Doel,$^{23}$
S.~Everett,$^{14}$
A.~E.~Evrard,$^{42,38}$
I.~Ferrero,$^{43}$
B.~Flaugher,$^{19}$
P.~Fosalba,$^{29,30}$
J.~Frieman,$^{19,8}$
J.~Garc\'ia-Bellido,$^{44}$
E.~Gaztanaga,$^{29,30}$
D.~W.~Gerdes,$^{42,38}$
T.~Giannantonio,$^{45,46}$
D.~Gruen,$^{47,11,24}$
R.~A.~Gruendl,$^{27,28}$
J.~Gschwend,$^{18,37}$
G.~Gutierrez,$^{19}$
W.~G.~Hartley,$^{48}$
S.~R.~Hinton,$^{49}$
D.~L.~Hollowood,$^{14}$
B.~Hoyle,$^{41,50}$
D.~Huterer,$^{38}$
D.~J.~James,$^{51}$
K.~Kuehn,$^{52,53}$
N.~Kuropatkin,$^{19}$
O.~Lahav,$^{23}$
M.~Lima,$^{17,18}$
M.~A.~G.~Maia,$^{18,37}$
M.~March,$^{12}$
J.~L.~Marshall,$^{54}$
F.~Menanteau,$^{27,28}$
J.~J.~Mohr,$^{41,50}$
R.~Morgan,$^{31}$
A.~Palmese,$^{19,8}$
F.~Paz-Chinch\'{o}n,$^{27,45}$
A.~Pieres,$^{18,37}$
A.~A.~Plazas~Malag\'on,$^{55}$
A.~Roodman,$^{11,24}$
E.~Sanchez,$^{39}$
V.~Scarpine,$^{19}$
M.~Schubnell,$^{38}$
S.~Serrano,$^{29,30}$
I.~Sevilla-Noarbe,$^{39}$
E.~Sheldon,$^{56}$
M.~Smith,$^{57}$
E.~Suchyta,$^{58}$
M.~E.~C.~Swanson,$^{27}$
G.~Tarle,$^{38}$
D.~Thomas,$^{59}$
C.~To,$^{47,11,24}$
T.~N.~Varga,$^{50,60}$
and J.~Weller$^{50,60}$
(DES Collaboration) 
\newline
\emph{\normalsize Affiliations are listed at the end of the paper}
}}

\date{Accepted XXX. Received YYY; in original form ZZZ}

\pubyear{2021}

\begin{document}
\label{firstpage}
\pagerange{\pageref{firstpage}--\pageref{lastpage}}
\maketitle

\begin{abstract}
The DMASS sample is a photometric sample from the DES Year 1 data set designed to replicate the properties of the CMASS sample from BOSS, in support of a joint analysis of DES and BOSS beyond the small overlapping area.  
In this paper, we present the measurement of galaxy-galaxy lensing using the DMASS sample as gravitational lenses in the DES Y1 imaging data.
We test a number of potential systematics that can bias the galaxy-galaxy lensing signal, including those from shear estimation, photometric redshifts, and observing conditions. After careful systematic tests, we obtain a highly significant detection of the galaxy-galaxy lensing signal, with total $S/N=25.7$. With the measured signal, we assess the feasibility of using DMASS as gravitational lenses equivalent to CMASS, by estimating the galaxy-matter cross-correlation coefficient $r_{\rm cc}$. By jointly fitting the galaxy-galaxy lensing measurement with the galaxy clustering measurement from CMASS, we obtain $r_{\rm cc}=1.09^{+0.12}_{-0.11}$ for the scale cut of $4 \hMpc$ and $r_{\rm cc}=1.06^{+0.13}_{-0.12}$ for $12 \hMpc$ in fixed cosmology. By adding the angular galaxy clustering of DMASS, we obtain $r_{\rm cc}=1.06\pm 0.10$ for the scale cut of $4 \hMpc$ and $r_{\rm cc}=1.03\pm 0.11$ for $12 \hMpc$. The resulting values of $r_{\rm cc}$ indicate that the lensing signal of DMASS is statistically consistent with the one that would have been measured if CMASS had populated the DES region within the given statistical uncertainty. The measurement of galaxy-galaxy lensing presented in this paper will serve as part of the data vector for the forthcoming cosmology analysis in preparation.
\end{abstract}


\input{intro_ggl}

\input{theory_ggl}

\input{data_ggl}

\input{measurement_ggl}

\input{result_ggl}

\input{conclusion_ggl}

\input{ack}

\section*{Data Availability}

The DMASS galaxy catalog used in this work and some of the ancillary data products shown in the Figures are available through the DES data release page (\url{https://des.ncsa.illinois.edu/releases/other}). Readers interested in comparing to these results are encouraged to check out the release page or contact the corresponding author for additional information.

\bibliography{dmass_ggl_reference,dmass_mg_reference}

\appendix

\section{ The impact of tails and a bump at \lowercase{$z\sim0.4$} in the redshift distribution of lens}
\label{sec:tails_bump}

\begin{figure*}
\centering
\includegraphics[width=1.0\textwidth]{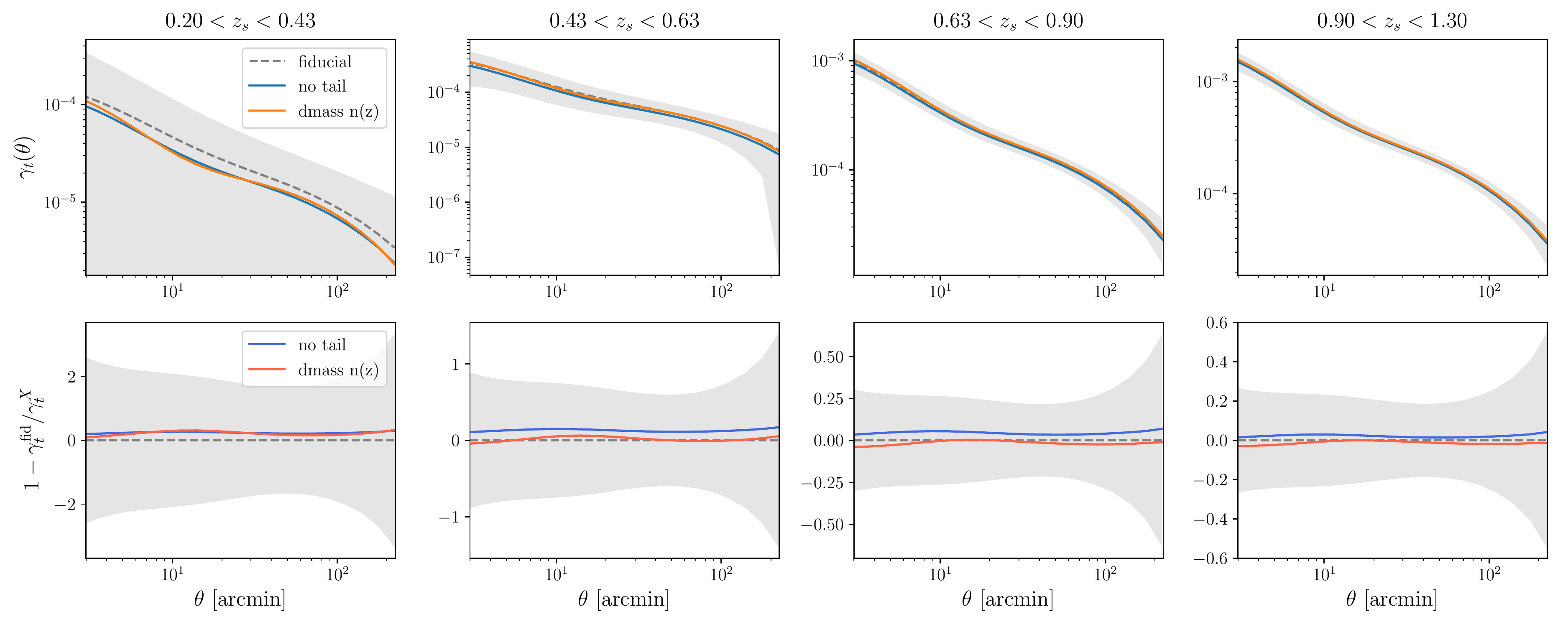}
\caption{ Theoretical prediction of tangential shear computed without low and high redshift tails (blue) and computed with the redshift distribution of DMASS including a bump at $z\sim0.4$ (red). The dashed line is computed with our fiducial setting. The predictions are well within the $1\sigma$ statistical error (shaded region) which implies the impacts from the tails and bump are negligible.  }
\label{fig:b_ggl_robustness}
\end{figure*}

The BOSS CMASS sample was selected by a set of photometric selection cuts before being targeted by the BOSS spectroscopy. Afterwards, the BOSS analyses only used sources within the redshift range of $0.43 < z < 0.75$, by applying the spectroscopic redshift cuts that discarded nearly $10\%$ of sources from the photometric targets \citep{Reid2016}. As the DMASS algorithm only replicates the photometric selection cuts, the resulting DMASS sample includes a small fraction of sources at the high-end ($z>0.75$) and low-end ($z < 0.43$). \cite{DMASS} tested the impact of these high- and low-redshift tails on the galaxy clustering using the photometric CMASS sample and found that the impact is negligible (see their appendix). However, for galaxy-galaxy lensing, the redshift tails of the lenses overlap with the redshift distributions of source bins, which might have a non-trivial impact.

To test the impact of the tails on galaxy-galaxy lensing, we compute the theoretical tangential shear using the spectroscopic redshift distribution of CMASS within $0.43 < z < 0.75$ and compare the result with the fiducial case computed with the full redshift distribution. The comparison with the fiducial case is shown in Figure \ref{fig:b_ggl_robustness}. The top row panels show the tangential shear with the full redshift distribution (`fiducial'; grey dashed) and the one with no redshift tails (`no-tails'; blue solid) for each source bin. The grey shaded area denotes the statistical uncertainty. The bottom row panels show the fractional difference between `fiducial' and `no-tail' (blue). The offset between `fiducial' and `no-tail' is within the statistical uncertainty. 

We also test the impact of the bump at the redshift $z \sim 0.4$ on galaxy-galaxy lensing. We compute the tangential shear signals with the clustering redshift distribution of DMASS (maroon color error bars in Figure \ref{fig:nz}) to take into account the bump and compare the resulting signals with the fiducial case, based on the spectroscopic redshift distribution of CMASS (red shaded region in Figure \ref{fig:nz}). The comparison with the fiducial case is shown in Figure \ref{fig:b_ggl_robustness} in orange color. We do not find any significant deviation from the fiducial case.

 \section*{Affiliations}
 \noindent
{\it
$^{1}$ Department of Physics, Duke University Durham, NC 27708, USA\\
$^{2}$ Center for Cosmology and Astro-Particle Physics, The Ohio State University, Columbus, OH 43210, USA\\
$^{3}$ Department of Physics, The Ohio State University, Columbus, OH 43210, USA\\
$^{4}$ Jet Propulsion Laboratory, California Institute of Technology, 4800 Oak Grove Dr., Pasadena, CA 91109, USA\\
$^{5}$ Department of Applied Mathematics and Theoretical Physics, University of Cambridge, Cambridge CB3 0WA, UK\\
$^{6}$ Department of Astronomy/Steward Observatory, University of Arizona, 933 North Cherry Avenue, Tucson, AZ 85721-0065, USA\\
$^{7}$ Department of Astronomy and Astrophysics, University of Chicago, Chicago, IL 60637, USA\\
$^{8}$ Kavli Institute for Cosmological Physics, University of Chicago, Chicago, IL 60637, USA\\
$^{9}$ Instituci\'o Catalana de Recerca i Estudis Avan\c{c}ats, E-08010 Barcelona, Spain\\
$^{10}$ Institut de F\'{\i}sica d'Altes Energies (IFAE), The Barcelona Institute of Science and Technology, Campus UAB, 08193 Bellaterra (Barcelona) Spain\\
$^{11}$ Kavli Institute for Particle Astrophysics \& Cosmology, P. O. Box 2450, Stanford University, Stanford, CA 94305, USA\\
$^{12}$ Department of Physics and Astronomy, University of Pennsylvania, Philadelphia, PA 19104, USA\\
$^{13}$ Department of Astronomy, University of California, Berkeley,  501 Campbell Hall, Berkeley, CA 94720, USA\\
$^{14}$ Santa Cruz Institute for Particle Physics, Santa Cruz, CA 95064, USA\\
$^{15}$ Department of Physics, Carnegie Mellon University, Pittsburgh, Pennsylvania 15312, USA\\
$^{16}$ Institute for Astronomy, University of Edinburgh, Edinburgh EH9 3HJ, UK\\
$^{17}$ Departamento de F\'isica Matem\'atica, Instituto de F\'isica, Universidade de S\~ao Paulo, CP 66318, S\~ao Paulo, SP, 05314-970, Brazil\\
$^{18}$ Laborat\'orio Interinstitucional de e-Astronomia - LIneA, Rua Gal. Jos\'e Cristino 77, Rio de Janeiro, RJ - 20921-400, Brazil\\
$^{19}$ Fermi National Accelerator Laboratory, P. O. Box 500, Batavia, IL 60510, USA\\
$^{20}$ Instituto de F\'{i}sica Te\'orica, Universidade Estadual Paulista, S\~ao Paulo, Brazil\\
$^{21}$ CNRS, UMR 7095, Institut d'Astrophysique de Paris, F-75014, Paris, France\\
$^{22}$ Sorbonne Universit\'es, UPMC Univ Paris 06, UMR 7095, Institut d'Astrophysique de Paris, F-75014, Paris, France\\
$^{23}$ Department of Physics \& Astronomy, University College London, Gower Street, London, WC1E 6BT, UK\\
$^{24}$ SLAC National Accelerator Laboratory, Menlo Park, CA 94025, USA\\
$^{25}$ Instituto de Astrofisica de Canarias, E-38205 La Laguna, Tenerife, Spain\\
$^{26}$ Universidad de La Laguna, Dpto. Astrofisica, E-38206 La Laguna, Tenerife, Spain\\
$^{27}$ Center for Astrophysical Surveys, National Center for Supercomputing Applications, 1205 West Clark St., Urbana, IL 61801, USA\\
$^{28}$ Department of Astronomy, University of Illinois at Urbana-Champaign, 1002 W. Green Street, Urbana, IL 61801, USA\\
$^{29}$ Institut d'Estudis Espacials de Catalunya (IEEC), 08034 Barcelona, Spain\\
$^{30}$ Institute of Space Sciences (ICE, CSIC),  Campus UAB, Carrer de Can Magrans, s/n,  08193 Barcelona, Spain\\
$^{31}$ Physics Department, 2320 Chamberlin Hall, University of Wisconsin-Madison, 1150 University Avenue Madison, WI  53706-1390\\
$^{32}$ Jodrell Bank Center for Astrophysics, School of Physics and Astronomy, University of Manchester, Oxford Road, Manchester, M13 9PL, UK\\
$^{33}$ University of Nottingham, School of Physics and Astronomy, Nottingham NG7 2RD, UK\\
$^{34}$ Astronomy Unit, Department of Physics, University of Trieste, via Tiepolo 11, I-34131 Trieste, Italy\\
$^{35}$ INAF-Osservatorio Astronomico di Trieste, via G. B. Tiepolo 11, I-34143 Trieste, Italy\\
$^{36}$ Institute for Fundamental Physics of the Universe, Via Beirut 2, 34014 Trieste, Italy\\
$^{37}$ Observat\'orio Nacional, Rua Gal. Jos\'e Cristino 77, Rio de Janeiro, RJ - 20921-400, Brazil\\
$^{38}$ Department of Physics, University of Michigan, Ann Arbor, MI 48109, USA\\
$^{39}$ Centro de Investigaciones Energ\'eticas, Medioambientales y Tecnol\'ogicas (CIEMAT), Madrid, Spain\\
$^{40}$ Department of Physics, IIT Hyderabad, Kandi, Telangana 502285, India\\
$^{41}$ Faculty of Physics, Ludwig-Maximilians-Universit\"at, Scheinerstr. 1, 81679 Munich, Germany\\
$^{42}$ Department of Astronomy, University of Michigan, Ann Arbor, MI 48109, USA\\
$^{43}$ Institute of Theoretical Astrophysics, University of Oslo. P.O. Box 1029 Blindern, NO-0315 Oslo, Norway\\
$^{44}$ Instituto de Fisica Teorica UAM/CSIC, Universidad Autonoma de Madrid, 28049 Madrid, Spain\\
$^{45}$ Institute of Astronomy, University of Cambridge, Madingley Road, Cambridge CB3 0HA, UK\\
$^{46}$ Kavli Institute for Cosmology, University of Cambridge, Madingley Road, Cambridge CB3 0HA, UK\\
$^{47}$ Department of Physics, Stanford University, 382 Via Pueblo Mall, Stanford, CA 94305, USA\\
$^{48}$ Department of Astronomy, University of Geneva, ch. d'\'Ecogia 16, CH-1290 Versoix, Switzerland\\
$^{49}$ School of Mathematics and Physics, University of Queensland,  Brisbane, QLD 4072, Australia\\
$^{50}$ Max Planck Institute for Extraterrestrial Physics, Giessenbachstrasse, 85748 Garching, Germany\\
$^{51}$ Center for Astrophysics $\vert$ Harvard \& Smithsonian, 60 Garden Street, Cambridge, MA 02138, USA\\
$^{52}$ Australian Astronomical Optics, Macquarie University, North Ryde, NSW 2113, Australia\\
$^{53}$ Lowell Observatory, 1400 Mars Hill Rd, Flagstaff, AZ 86001, USA\\
$^{54}$ George P. and Cynthia Woods Mitchell Institute for Fundamental Physics and Astronomy, and Department of Physics and Astronomy, Texas A\&M University, College Station, TX 77843,  USA\\
$^{55}$ Department of Astrophysical Sciences, Princeton University, Peyton Hall, Princeton, NJ 08544, USA\\
$^{56}$ Brookhaven National Laboratory, Bldg 510, Upton, NY 11973, USA\\
$^{57}$ School of Physics and Astronomy, University of Southampton,  Southampton, SO17 1BJ, UK\\
$^{58}$ Computer Science and Mathematics Division, Oak Ridge National Laboratory, Oak Ridge, TN 37831\\
$^{59}$ Institute of Cosmology and Gravitation, University of Portsmouth, Portsmouth, PO1 3FX, UK\\
$^{60}$ Universit\"ats-Sternwarte, Fakult\"at f\"ur Physik, Ludwig-Maximilians Universit\"at M\"unchen, Scheinerstr. 1, 81679 M\"unchen, Germany\\

}

\bsp   
\label{lastpage}
\end{document}

%% file: intro_ggl.tex
\section{Introduction}
\label{sec:intro}

Galaxies are biased density tracers as they form at the peaks of the matter density field \citep{Kaiser1984}. To interpret the observed distribution of galaxies accurately, one needs to understand the relation between the galaxy and matter density fields. At large scales, the galaxy density field is proportional to the matter density. The ratio between the galaxy and matter clusterings can be related by a constant factor, often referred to as linear galaxy bias. 
On small scales, non-linearity and stochasticity induce more complexity in the relation, making the modeling of the correlations between two fields more challenging \citep{Tegmark1998, Dekel1999,Tegmark1999}.  

The relationship between galaxy and underlying matter distribution can be studied using other means, such as galaxy-galaxy lensing. 
Galaxy-galaxy lensing uses the subtle distortion of background galaxy shapes to infer the mass profile surrounding foreground galaxies. Under the linear assumption, the strength of the galaxy-galaxy lensing signal depends on the product of galaxy bias and the amplitude of matter clustering ($\propto b \sigma_8^2$), while the amplitude of galaxy clustering depends on the galaxy bias squared ($\propto b^2 \sigma_8^2$). 
Hence, the combination of the two probes yields a high precision measurement of the amplitude of matter clustering, by cancelling out galaxy bias that has been a major source of uncertainty in cosmological analyses (see, e.g., \citepp{Yoo2012} and \citepp{Park2016CombiningClustering}).

The spectroscopic galaxy samples from the Baryon Oscillation Spectroscopic Survey \citep[BOSS;][]{Dawson2013TheSDSS-III}, referred to as `LOWZ' and `CMASS' \citep{Reid2016}, yielded the most precise measurements of baryonic acoustic oscillations (BAO) and redshift space distortions (RSD) from the full shape of the galaxy correlation function in the redshift range of $0.1 < z < 1.0$ \citep{Alam2017}. Due to the large sample size and the availability of spectroscopic redshifts, the two samples have also been a popular candidate for gravitational lenses, to optimally combine the weak lensing signals from background sources with the galaxy clustering measurements of the BOSS galaxies.
Several studies \citep{Miyatake2015, More2015, Alam2017TestingCMASS, Singh2020, Amon2018, Jullo2019} have conducted a joint analysis of galaxy clustering and weak lensing using the BOSS galaxies as gravitational lenses on the deep imaging data from modern experiments, such as the Canada-France-Hawaii Telescope Lensing Survey \citep[CFHTLenS;][]{CFHTLenS} and Kilo-Degree Survey \citep[KiDS;][]{KIDS}. This approach provides access to better deep images while maintaining the strong constraining power of the galaxy clustering measurement from BOSS. However, the lensing measurements of these analyses are restricted to a small overlapping area between BOSS and imaging surveys, mostly within only a few hundreds of ${\rm deg}^2$.

The Dark Energy Survey \citep[DES;][]{DESOverview} is a prime candidate for such an analysis for its precise photometry and the largest survey area among the current generation of Stage-III experiments. 
The survey images over $5,000\degsqr$ of the southern sky in the $grizY$ bands for a wide-area survey and $27\degsqr$ `time domain' fields in the $griz$ bands for supernovae. 
Despite the most extensive survey area among the modern experiments, the overlapping region between the DES Year 1 footprint ($\sim1,800\degsqr)$ and the BOSS footprint is fairly small, consisting of only \stripeGold, comparable to  previous measurements combining lensing and clustering.  

To overcome the aforementioned limitations, \cite{DMASS} constructed a probabilistic model that identifies galaxies equivalent to the BOSS CMASS galaxies in the DES footprint, extending beyond the overlapping region. The resulting galaxy sample, DES-CMASS (hereafter DMASS), covers the lower region of the DES wide-area survey footprint scanned during the first-year observations of DES ($1,244\deg^2$), which effectively increases the area available for such studies by a factor of 10. Through a series of validation tests, \cite{DMASS} showed that DMASS has the same properties as the BOSS CMASS sample, such as the galaxy number density, redshift distribution, and angular galaxy clustering. 

This paper has two specific goals. 
First, we measure the galaxy-galaxy lensing signal using the DMASS sample as gravitational lenses on the DES Y1 imaging data. The measured signals are calibrated by removing contamination from various systematics and astrophysical effects.  
Second, using the calibrated measurement, we assess the feasibility of using DMASS as gravitational lenses equivalent to CMASS. For this, we quantify the difference in galaxy bias from galaxy-galaxy lensing of DMASS and galaxy clustering of BOSS CMASS, by estimating the galaxy-matter cross-correlation coefficient $r_{\rm cc}$ in the scales of interest
(see, e.g., \citepp{Schneider1998}, \citepp{vanWaerbeke1998}, \citepp{Hoekstra2001}, \citepp{Hoekstra2002}, \citepp{Baldauf2010}, \citepp{Prat2018DarkLensing} and \citepp{Simon2021} for classical and recent works of $r_{\rm cc}$).
On large scales where the linear bias assumption is valid, the matter density field and the galaxy density field are fully correlated such that $r_{\rm cc}$ approaches unity.  
In this work, the value of $r_{\rm cc}$ equal to one implies that the galaxy-galaxy lensing signal of DMASS on large scales where the linear theory is valid can be considered as being statistically consistent with the one that would have been measured if CMASS populated the full DES region. The lensing signals presented in this work will be utilized as part of the data vectors for a combined analysis of BOSS and DES in a forthcoming work \citep{DMASS-MG}.

This paper is organized as follows. In Section \ref{sec:theory}, we introduce the theory of weak lensing and the cross-correlation coefficient $r_{\rm cc}$. The data sets used in the analysis are described in Section \ref{sec:data}. Models, parameters, and other analysis choices can be found in Section \ref{sec:measurement}. In Section \ref{sec:result}, we present our estimates of galaxy bias and the galaxy-matter cross-correlation coefficient.
Conclusions and discussions are presented in Section \ref{sec:conclusion}.

The fiducial cosmological model used throughout this paper is the {\it Planck} 2018 cosmology \citep{Planck2018CosmologicalParameter} 
with the following parameters: matter density $\Omega_{m} = 0.315$, baryon density $\Omega_b = 0.049$, 
amplitude of matter clustering $\sigma_8 = 0.815$, spectral index $n_s = 0.965$ and Hubble constant $h \equiv H_0/100~{\rm{km~ s^{-1} Mpc^{-1}}} = 0.674$. 
Our choice for the fiducial cosmology does not affect the measurement of the cross-correlation coefficient $r_{\rm cc}$. This is because the measurement of CMASS galaxy clustering \citep{Chuang2017} used in this work is consistent with the {\it Planck} 2018 cosmology, and the quantity $r_{\rm cc}$ depends on the relative difference in the amplitude of galaxy clustering and galaxy-galaxy lensing.

%% file: theory_ggl.tex
\section{Theory}
\label{sec:theory}
Weak gravitational lensing is the deflection of light from distant objects by foreground matter in the Universe. 
In the case of galaxy lensing, the lensing effect distorts the shapes of background galaxies.
Since light from distant sources must pass by nearby foreground matter distributions, the distortion can inform us about the distribution of matter in between the source and us (for a detailed review, see \citepp{Huterer225}).

The distortions of images of background galaxies can be described as 
\bea
\left( 
\begin{matrix}
	x_{\rm u} \\
	y_{\rm u} \\
\end{matrix}
\right)
=
\left(
\begin{matrix}
	1-\kappa - \gamma_1 & -\gamma_2 \\
	-\gamma_2 & 1-\kappa+\gamma_1 \\
\end{matrix}
\right)
\left(
\begin{matrix}
	x_{\rm l} \\
	y_{\rm l} \\
\end{matrix}
\right)
\eea
where $(x_{\rm u}, y_{\rm u})$ is the displacement vector in the source plane and $(x_{\rm l}, y_{\rm l})$ is the displacement vector in the image plane. The subscripts `u' and `l' denote `unlensed' and `lensed' respectively. $\gamma_1$ and $\gamma_2$ are the real and imaginary components of the total lensing shear $\gamma$.
The total lensing shear is defined as $\gamma = \gamma_1 + i\gamma_2$. 

The main observable for measuring galaxy-galaxy lensing is the tangential shear of background sources relative to the line joining the lens and source. For a given lens-source pair, the equations for the tangential shear and cross-components of the shear are given by 
\bea
\gamma_{+} &=& - {\rm Re} [ \gamma e^{-2 i \phi} ]~, \\ 
\gamma_{\times} &=& - {\rm Im} [ \gamma e^{-2 i \phi} ]~,
\eea
where $\phi$ is the position angle of the source galaxy with respect to the horizontal axis of the Cartesian coordinate system centered at the lens. The signal of the shear is typically very subtle compared to the intrinsic ellipticity of a source galaxy. To obtain an estimate of shear with a significant signal-to-noise ratio, one needs to average over many galaxies. Hence, the ensemble average of the tangential shear is conveniently used as the theoretical expression for galaxy-galaxy lensing, which is defined as
\bea
\gamma_{\rm t}(\theta) = \langle \gamma_{+} (\theta) \rangle~,
\eea 
at an angular separation $\theta$. 
The mean tangential shear $\gamma_{\rm t}(\theta)$ can be expressed as the Fourier transform of the galaxy-matter angular power spectrum $C_{\rm g\kappa}$ as below:
\bea
\gamma_{\rm t} (\theta) = \frac{1}{2\pi} \int^{\infty}_{0} C_{\rm g \kappa} (\ell) J_{2} ( \ell \theta) l \ud \ell  ~,
\label{eq:gammat}
\eea
where $l$ denotes the angular multipole, $J_2(x)$ is the second order Bessel function of the first kind. The galaxy-matter angular power spectrum $C_{\rm g\kappa}$ is the projection along the line of sight of the 3D power spectrum as given by \citep{Kaiser1992,LoVerde2008}
\bea
C_{\rm g \kappa} (\ell) = \int^{\infty}_{0} \ud \chi \frac{W_{\rm g} ( k, \chi) W_{\kappa} (\chi) }{\chi^2} P_{\rm g\delta} (k, z(\chi))~.
\eea 
where $\chi$ is the comoving distance, $k=(\ell+1/2)/\chi$ under the Limber approximation, and $P_{\rm g\delta}(k, z(\chi))$ is the galaxy-matter cross-power spectrum.  
The integral along the line of sight indicates that weak lensing radially projects the density fluctuations between us and the source galaxies. 
The function $W_{\kappa}(\chi)$ is the geometric weight function describing the lensing efficiency defined as
\bea
W_{\kappa}(\chi) = \frac{3 H_0^2 \Omega_{\rm m}}{2c^2}  \frac{\chi}{a(\chi)}  \int^{\infty}_{\chi} \ud \chi' \frac{n_{\kappa}(z(\chi') ) \ud z/\ud \chi' }{ \bar{n}_{\kappa} } \frac{\chi' - \chi}{\chi'}~  
\eea
in terms of the source distribution $n_{\kappa}(\chi')$. The quantity $\bar{n}_{\kappa}$ is the number density for sources defined as $\bar{n}_{\kappa} = \int \ud z ~ n_{\kappa} (z)$. 
The function $W_{\rm g}(\chi)$ is the geometric weight function for clustering given as 
\bea
W_{\rm g} (k,\chi) = \frac{n_{\rm g} (z(\chi)) }{\bar{n}_{\rm g}} \frac{\ud z}{\ud \chi}~,
\eea
where 
$n_{\rm g}$ is the redshift distribution of the lens galaxies, and $\bar{n}_{\rm g}$ is the number density for lenses. 
In the regime where the linear relationship between the galaxy and matter densities holds, 
the galaxy-matter cross-power spectrum $P_{\rm g\delta}$ is defined as a combination of the nonlinear matter power spectrum ($P_{\delta \delta}$) and galaxy bias ($b$) as follows: $P_{\rm g\delta}(k,z(\chi)) = b(k,z(\chi)) P_{\delta \delta}(k,z(\chi))$. 
However in the weakly nonlinear regime at scales of a few $\hMpc$, nonlinear effects and stochasticity between matter and galaxy densities may result in the two fields being less correlated. 
Hence, to incorporate the correlation relationship between two fields, $P_{\rm g\delta}$ is defined as \citep{Pen1998, Tegmark1999}
\bea
P_{\rm g\delta}(k, z(\chi)) = b(k, z(\chi)) \, r_{\rm cc} (k, z(\chi))\,  P_{\delta \delta} (k,z(\chi))~, 
\eea
with the correlation coefficient $r_{\rm cc}$ defined as 
\bea
r_{\rm cc} (k, z(\chi)) = \frac{ P_{\rm g\delta} (k, z(\chi)) }{ \sqrt{P_{\rm gg}(k, z(\chi)) \, P_{\delta \delta}(k, z(\chi)) } }~,
\eea
where $P_{\rm gg}$ is the galaxy power spectrum.  
The relation between the galaxy power spectrum and the matter power spectrum is given as $P_{\rm gg}(k,z(\chi)) = b(k,z(\chi))^2 P_{\delta \delta} (k,z(\chi))$, which remains unchanged. 
%
On large scales where the linear bias assumption is valid, the matter density field and the galaxy density field are fully correlated such that the correlation coefficient $r_{\rm cc}$ approaches unity \citep{Dekel1999,Somerville2001, Baldauf2010}. 
Assuming that the galaxy bias is weakly dependent on scales and redshift in our lens sample, the combination of $b$ and $r_{\rm cc}$ can be taken out of the integrals as below
\bea
C_{\rm g \kappa}(\ell) =  b\, r_{\rm cc}  \int^{\infty}_{0} \ud \chi  \frac{W_{\rm g}(k, \chi) W_{\kappa} (\chi) }{\chi^2} P_{\delta\delta} (k, z(\chi))~,
\label{eq:cgk}
\eea 
where $b\, r_{\rm cc}$ is an averaged quantity over the redshift range of the lens bin. Then, the tangential shear $\gamma_{\rm t}$ is simply proportional to $b\, r_{\rm cc} \sigma_8^2$.
  
In this work, we will mainly use scales where the linear bias model is valid, 
and obtain the measurement of 
$r_{\rm cc}$ from the combination of galaxy-galaxy lensing from DMASS and galaxy clustering from BOSS CMASS fixing the cosmology to that of  {\it Planck} 2018 \citep{Planck2018CosmologicalParameter}. 
We define the galaxy bias constraint inferred from galaxy-galaxy lensing as $b_{\gamma} = b\, r_{\rm cc}$. 
Then, the value of $r_{\rm cc}$ can be derived from the ratio of the two galaxy bias constraints given as
\bea
b=b_{\rm g}~;~~r_{\rm cc} = \frac{b_{\gamma}}{b_{\rm g}}~,
\eea
where $b_{\rm g}$ represents the linear galaxy bias from galaxy clustering of BOSS CMASS. 
As we use different tracers for galaxy clustering and galaxy-galaxy lensing, 
the measurement of $r_{\rm cc}$ in our work not only shows the cross-correlation between galaxies and matter within the scales of interest, but it can also be interpreted as a barometer indicating the consistency between the two tracers. 


%% file: data_ggl.tex
\section{Data}
\label{sec:data}

\begin{figure}
\centering
\includegraphics[width=0.45\textwidth]{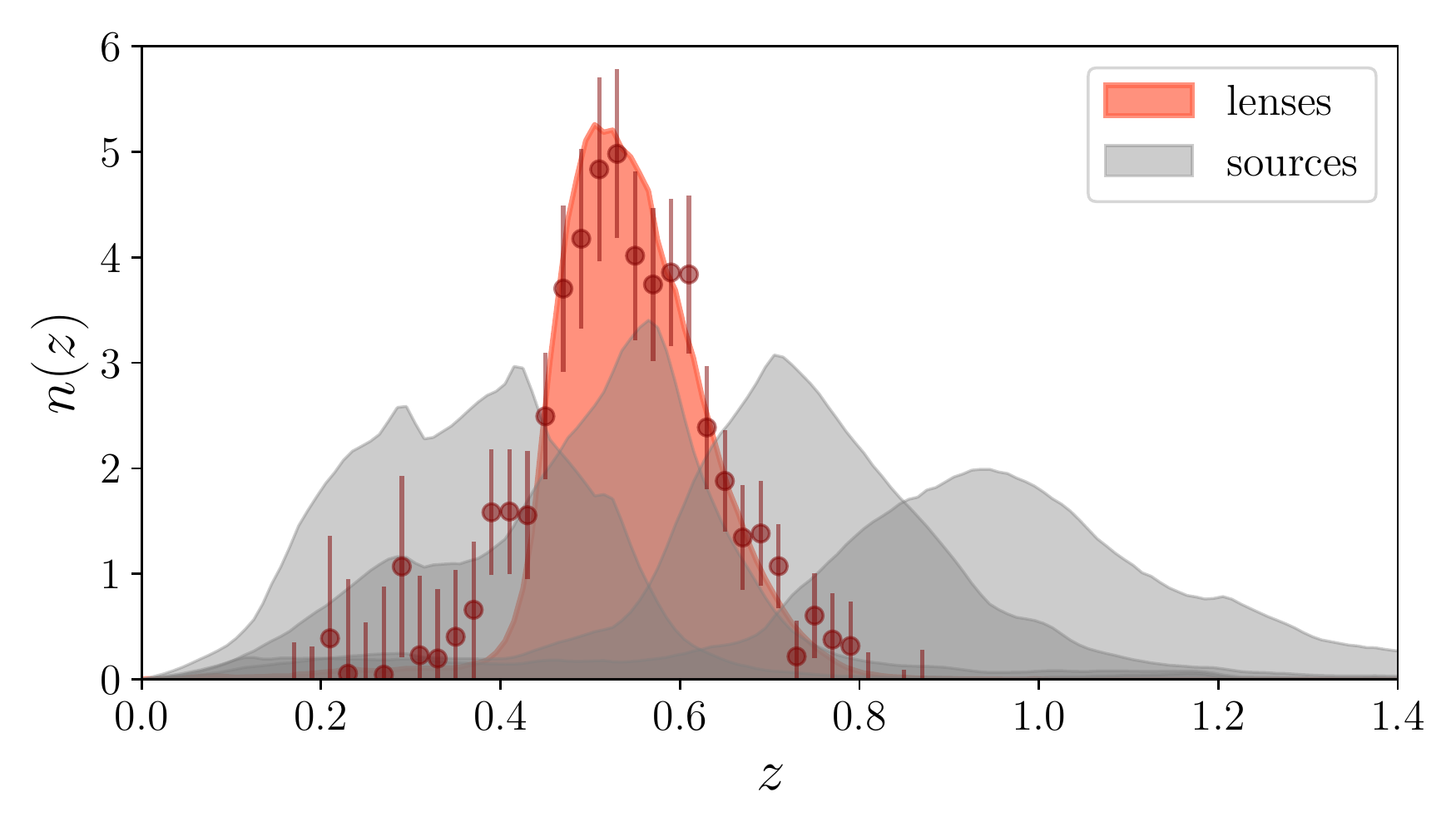}
\caption{ 
Redshift distributions of lenses (red) and sources (grey) used for theoretical predictions. 
In this work, we adopt the spectroscopic redshift distribution of CMASS for lenses, as the redshift distribution of DMASS obtained from the clustering-$z$ method shows a good agreement with CMASS. 
The redshift distribution of DMASS is over-plotted in maroon color with error bars. 
The source sample from DES Y1 \metacal\ is divided into 4 tomographic bins  ($0.2 < z < 0.43$, $0.43 < z < 0.63$,  $0.63 < z < 0.90$ and $0.90 < z < 1.30$) using the mean of the photo-$z$ probability density function determined with the BPZ photometric redshift code. 
}
\label{fig:nz}
\end{figure}

In this section, we describe the data sets we use for the analysis. 
For the galaxy clustering measurement, we utilize the RSD and BAO measurements from the BOSS CMASS galaxy sample \citep{Chuang2017}.
For the galaxy-galaxy lensing measurement, we use the DMASS galaxy catalog \citep{DMASS} and \metacal~ shape catalog \citep*{METACAL1,METACAL2,Zuntz2018} from DES. 
Both catalogs are based on the images taken between Aug. 31, 2013 and Feb. 9, 2014 during the first-year observations of DES \citep{DESCollaboration2006, Flaugher2015THECAMERA, DESCollaboration2017}. The scanned area during the period is about $1514\deg^2$ after masking bad regions with 
a $10\sigma$ limiting magnitude of $i=22.5$ for galaxies \citep{Y1GOLD}. 
Below, we briefly describe these data sets and refer readers to the listed references for more details. 

\subsection{Galaxy Clustering: BAO and RSD measurements from BOSS CMASS}
\label{sec:data.cmass}

The Baryon Oscillation Spectroscopic Survey 
\citep[BOSS;][]{Eisenstein2011BOSS, Bolton2012SpectralSurvey, Dawson2013TheSDSS-III} was designed to measure the characteristic scale imprinted by baryon acoustic oscillations (BAO) with a precision of $\sim1\%$, over a larger volume than the combined efforts of all previous spectroscopic surveys. 
BOSS targeted two distinct samples known as LOWZ at $0.15 < z < 0.43$ and CMASS at $0.43 < z < 0.75$. 
The higher redshift sample, CMASS, we focus on in this work was designed to select a stellar mass-limited sample of objects of all intrinsic colors, with a color cut that selects almost exclusively on redshift. 
Every galaxy satisfying the selection cuts was targeted by the BOSS spectrograph to obtain their spectroscopic redshifts, except for $5.8\%$ of galaxies in a fiber collision group and $1.8\%$ of galaxies for which the spectroscopic pipeline fails to obtain a robust redshift \citep{Reid2016}.
%
\cite{Chuang2017} presented the constraints of BAO and RSD derived from 
galaxy clustering of the combined BOSS galaxy samples. They provided a set of values of the Hubble parameter ($H(z)$), the angular diameter distance ($d_{\rm A}(z)$), the matter density fraction ($\Omega_{\rm m} h^2$), the linear growth rate  and mean galaxy bias combined with the amplitude of mass fluctuation ($f(z) \sigma_8(z)$, $b \sigma_8(z)$) along with covariances between those parameters. In this work, we utilize those constraints measured at the mean redshift of CMASS ($z=0.59$) and the corresponding covariance matrix.

\subsection{Galaxy-galaxy lensing}
 \subsubsection{Lenses: DMASS}
 \label{sec:data.lens}
 
The DMASS galaxy sample is a subset of the DES Gold catalog, which consists of $\sim 137$ million clean objects validated for accurate cosmological analyses \citep{Y1GOLD}.
The sample was specifically designed to replicate the statistical properties of the BOSS CMASS sample \citep{Reid2016}, in support of upcoming joint analyses of the weak lensing measurements from DES and the existing measurements of galaxy clustering from BOSS. The sample selection algorithm was trained and validated by the DES photometry from the overlapping area between the DES and BOSS footprints.
The final selected sample consists of $117,293$ effective galaxies covering $1,244 \degsqr$ after masking bad regions described in \cite{DMASS}. The mean galaxy bias constrained by its angular galaxy clustering achieved $1\sigma$ consistency with the mean galaxy bias from the angular galaxy clustering of CMASS. The redshift distribution of DMASS was estimated by cross-correlating with the DES Y1 \redmagic galaxy sample \citep{ELVINPOOLE} and showed a good agreement with the spectroscopic redshift distribution of CMASS. 
The redshift distributions of CMASS (red shaded region) and DMASS (maroon error bars) are shown in Figure \ref{fig:nz}. 
The impact of the bump at $z\sim0.4$ on galaxy-galaxy lensing is found to be negligible as described in Appendix \ref{sec:tails_bump}. 
Hence, we adopt the spectroscopic redshift distribution of CMASS\footnote{The BOSS analyses use the CMASS galaxies only within the redshift range $( 0.43 < z < 0.75 )$, by applying the spectroscopic redshift cuts on the CMASS targets selected by a set of photometric cuts \citep{Reid2016}. 
However, we do not remove the low- and high-end redshift tails because the DMASS algorithm only replicates the photometric selection cuts of CMASS. Therefore, the resulting DMASS sample includes a small fraction of sources at the tails as well. \cite{DMASS} tested the impact of the redshift tails on the galaxy clustering of BOSS CMASS and found that the impact is negligible. The impact of the redshift tails on galaxy-galaxy lensing is described in Appendix \ref{sec:tails_bump}.} 
as a true redshift distribution of DMASS for theoretical predictions. 
For further details of the galaxy sample and selection algorithm, we refer readers to \cite{DMASS}. 

\subsubsection{Sources: DES Y1  \metacal}
\label{sec:data.source}

We adopt the \metacal\ catalog as sources. 
\metacal\ in the catalog name refers to a method to calibrate the bias in shear estimation by artificially shearing the galaxy images and re-measuring the shape \citep{METACAL1, METACAL2}.
As in \citet*{Zuntz2018, Prat2018DarkLensing} and \cite{Troxel2018}, we only keep clean sources with flag \verb|FLAGS_SELECT = 0| and split the sources into four tomographic bins by the mean photo-$z$ between $z=0.2$ and $z=1.3$.  
Photo-$z$ of individual galaxies are estimated by the Bayesian Photometric Redshift (BPZ) algorithm \citep{BPZ}. Further descriptions of the photo-$z$ catalog associated with the shear catalogs can be found in \citet*{Hoyle2018}. 
The shear multiplicative biases, photo-$z$ biases, and their uncertainties related to this catalog are quantified in \citet*{Zuntz2018} and \citet*{Hoyle2018} and employed as priors in our analysis. See Section \ref{sec:measurement.likelihood} for a detailed description.

%% file: measurement_ggl.tex
\section{Measurement}
\label{sec:measurement}

In this section, we describe our methodology of measuring the mean tangential shear $\gamma_{\rm t}$ in configuration space using the DMASS and \metacal~catalogs.  
The estimator for measuring tangential shear is explained in Section \ref{sec:measurement.Estimator}. 
In Section \ref{sec:measurement.scales}, we select scales where the linear bias model is valid based on analyses performed in the past. 
In Section \ref{sec:measurement.CovarianceMatrix}, we compute the theoretical covariance matrix and validate it with the jackknife method. 
In Section \ref{sec:measurement.Systematics.boost}, we calculate boost factors. 
The impact of various systematics and astrophysical effects are outlined in Section \ref{sec:measurement.Systematics}.  
Finally, in Section \ref{sec:measurement.likelihood}, we measure the cross-correlation coefficient $r_{\rm cc}$ by combining the resulting $\gamma_{\rm t}$ and the measurements of galaxy clustering from BOSS CMASS fixing the cosmology to that of  {\it Planck} 2018.

For the purposes of measuring $\gamma_{\rm t}$, 
we use 
four source bins selected using BPZ: $0.2 < z < 0.43 $, $0.43 < z < 0.63 $, $0.63 < z < 0.90 $ and $0.90 < z < 1.30 $ as shown in \citet*{Prat2018DarkLensing}. We do not divide the lens sample. 
The redshift distributions of the lens and four source bins are shown in Figure \ref{fig:nz}. 
The weights and masks for removing systematics in lenses are addressed in Section 4 in \cite{DMASS}. For the systematic characterization for the source bins, see \citet*{Prat2018DarkLensing} and \cite{Troxel2018}. 
All calculations of correlation functions are performed in 20 logarithmically spaced angular bins over the range $2.5 \arcmin < \theta < 250 \arcmin$ using the public code \verb|TreeCorr|\footnote{\url{https://github.com/rmjarvis/TreeCorr}}\citep{TreeCorr}.
For all of our measurements, 
we use jackknife (JK) resampling \citep{Norberg2009}. The survey area is split into HEALPix\footnote{\url{http://healpix.sourceforge.net}} \citep{HEALPix} pixels at resolution $N_{\rm side} = 16$. 
This results in $\sim 170$ jackknife regions of $\sim 13 \degsqr$, comparable to the maximum angular scales of $250 \arcmin$. We find that the impact of the unequal size of pixels at the edge of the footprint is negligible as $\sim 80$ jackknife patches generated by the 
\verb+kmeans+\footnote{\url{https://github.com/esheldon/kmeans_radec}}   
 algorithm yield a consistent result. 

\subsection{Estimator}
\label{sec:measurement.Estimator}
We measure the mean tangential shear by averaging over many lens-source pairs as below:
\bea
\gamma_{\rm t}(\theta) = \langle \gamma_{+} (\theta) \rangle =\frac{1}{\langle R \rangle } \frac{\sum_{j} w_{ {\rm ls},j} \gamma_{+,j} }{\sum_{j} w_{{\rm ls},j} }~,
\label{eq:gamma}
\eea
where the subscripts `${\rm l}$' and `${\rm s}$' denote lenses and sources. The notation $w_{\rm ls}$ is a combination of weights associated with each lens-source pair given as 
\bea
w_{\rm ls} = w_{\rm dmass} w_{\rm sys} ~,
\label{eq:weight}
\eea
where $w_{\rm dmass}$ is the probability of a galaxy being a member of the DMASS sample, $w_{\rm sys}$ is a weight for lens galaxies to correct the systematics due to observing conditions \citep{DMASS}. 
%
The value $\langle R \rangle$ in the denominator is the mean shear response averaged over the sources, which is defined as the sum of the measured shear response ($R_{\gamma}$) and shear selection bias  correction matrix ($R_{\rm S}$) for \metacal ~as below:
\bea
\langle R \rangle = \langle R_{\gamma} \rangle + \langle R_{\rm S} \rangle ~.
\eea
Finally, to remove additive systematics arising due to the survey edge or heavily masked regions, the signal around random points is subtracted from the signal around lens galaxies as below \citep[e.g.,][]{Mandelbaum2005,Mandelbaum2013,Singh2017}:
\bea
\gamma_t(\theta) =  \gamma_t^{\rm lens}(\theta) -  \gamma_t^{\rm random}(\theta)~.
\eea
Random points are uniformly generated on the surface of a sphere and masked by the same veto masks applied to the lens sample. The number density of randoms is chosen to be $50$ times denser than the lens sample, minimizing the impact of any noise from the finite number of randoms.

\subsection{Scale cuts}
\label{sec:measurement.scales}

The assumption of linear galaxy bias is expected to break down at small scales. Therefore, we try to restrict our analysis to sufficiently large scales where our modeling is valid. 
\cite{Baldauf2010} suggested removing the small scale information that is strongly affected by the stochastic relation between galaxies and matter. They found $r_{\rm cc} \sim 1$ using the comoving scales $r > 2 r_{\rm vir}$, where $r_{\rm vir}$ is the virial radius of haloes in the sample. 
Following the approach of \cite{Baldauf2010}, \cite{Singh2020} developed the methodology to constrain the cosmological parameters from the combination of galaxy clustering and galaxy-lensing cross-correlations. They modeled the galaxy-matter cross-correlation coefficient using the mock catalogs of BOSS CMASS and LOWZ galaxies and confirmed  $r_{\rm cc}$ to be consistent with unity above the cut-off scale of $\sim 2 \hMpc (> 2 r_{\rm vir})$. 
\cite{More2015} estimated galaxy bias and $r_{\rm cc}$ as a function of scales by combining the clustering and the galaxy-galaxy lensing signal of CMASS galaxies on the CFHTLenS images. 
The measured quantity of $r_{\rm cc}$ shows significant deviations from unity at small scales while being unity on the scales of $r > 10 \hMpc$. Similarly, \cite{Alam2017TestingCMASS} investigated the impact of nonlinearity at small scales utilizing galaxy clustering and galaxy-galaxy lensing of CMASS galaxies predicted from $N$-body simulations. In their work, the impact of nonlinearity on galaxy bias has a maximum value at $8 \hMpc$ and approaches nearly zero at $\sim 12 \hMpc$. 
Based on these aforementioned works, 
we choose a comoving scale cut of $12 \hMpc$ as our fiducial cut, and compare the result with a more aggressive scale cut of $4 \hMpc$.

The angular scale cut corresponding to the given comoving scale cut $r_{\rm min}$ is calculated as 
\bea
\theta_{\rm min} = \frac{ r_{\rm min} }{ \chi ( \langle z \rangle)}~,
\eea
where $\langle z \rangle = 0.59$ is the mean redshift of the DMASS sample. 
Hence, the corresponding angular scale cuts for $4 \hMpc$ and $12 \hMpc$  are obtained as $9\arcmin$ and $27\arcmin$, respectively.

\subsection{Covariance matrix}
\label{sec:measurement.CovarianceMatrix}

\begin{figure}
\centering
\includegraphics[width=0.45\textwidth]{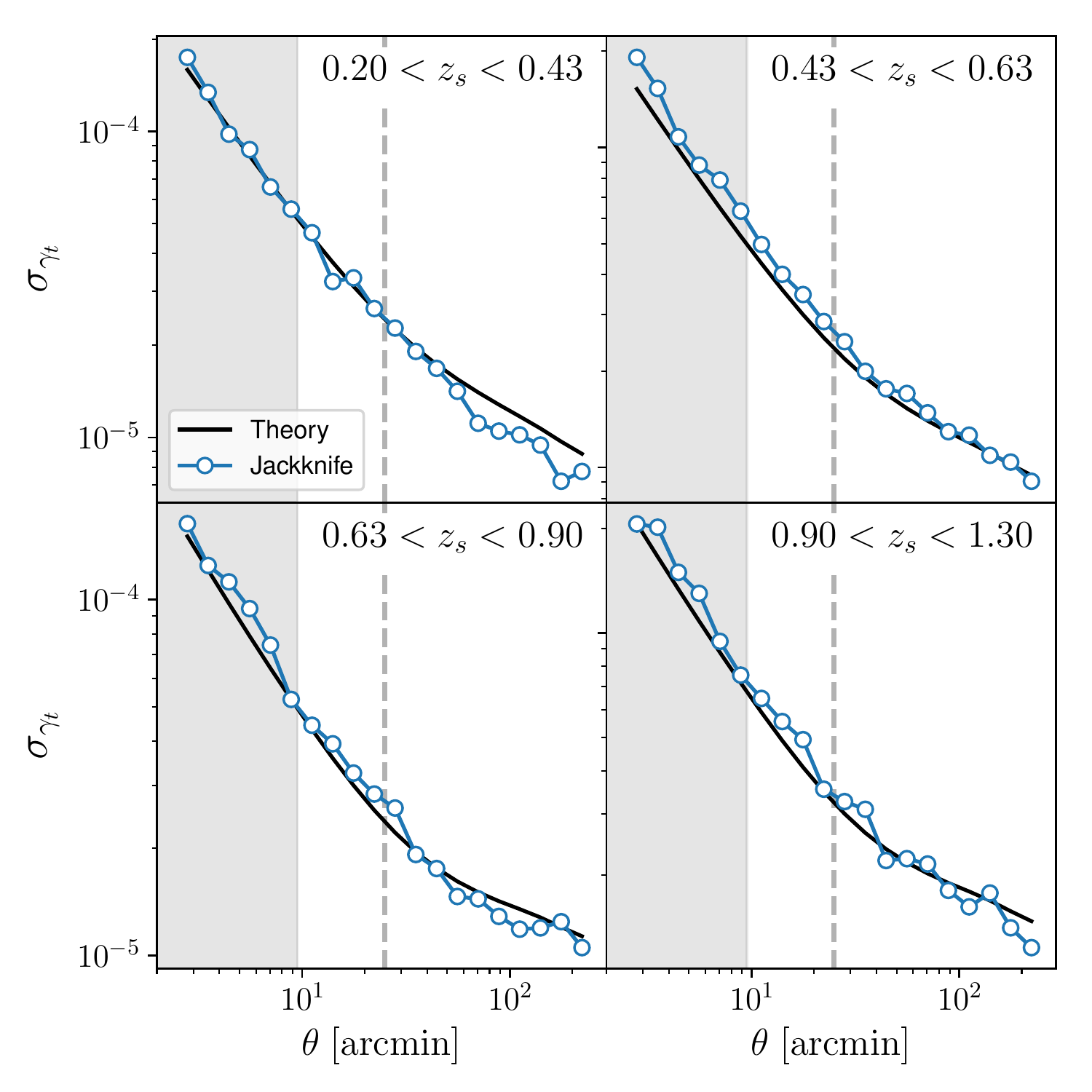}
\caption{Comparison of the diagonal components of the covariance obtained from theoretical computation (black solid) and the jackknife method on the data (blue circle), for all the lens-source combinations. The shaded region indicate the small scales that are removed by the $4 \hMpc$ scale cut. The vertical dashed lines indicate the scale cut of $12 \hMpc$. 
The overall amplitudes exhibit a good agreement with theory over the scales of interest ($>4\hMpc$). 
}
\label{fig:cov-diagonal}
\end{figure}

We obtain the statistical uncertainties of galaxy-galaxy lensing from a covariance matrix calculated by \cosmolike~\citep{COSMOLIKE}. The covariance is computed as the sum of Gaussian covariance and non-Gaussian covariance, and the super-sample covariance as detailed in \cite{Krause2017}. 
%

To validate the theoretical covariance matrix, we compare it with one computed by the jackknife (JK) method as below:
\bea
{\bf{C}}(\gamma_{i}, \gamma_{j}) = \frac{1}{N_{\rm JK} - 1} \sum^{N_{\rm JK}}_{k=0} (\gamma_{i}^k - \bar{\gamma}_{i} ) (\gamma_{j}^k -\bar{\gamma}_{j}) ~,
\eea
where $N_{\rm JK}$ is the total number of JK samples, $\gamma_i$ represents the $i$th bin of the tangential shear, $\gamma_i^{k}$ denotes the $i$th bin of the tangential shear from the $k$th sample, and $\bar{\gamma}$ is the average value of $\gamma$ over all samples. 
The footprint is split into HEALPix pixels at resolution $N_{\rm side} = 16$ that results in $176$ JK sub regions. 
In order to correct a biased estimate of the inverse covariance, the Hartlap correction factor $ (N_{\rm JK} - N_{\rm bins}-2)/(N_{\rm JK} - 1)$ is applied, where $N_{\rm bins}$ is the number of angular bins \citep{Hartlap2007}.

The four panels in Figure \ref{fig:cov-diagonal} display the diagonal components of correlation matrices calculated from theory (black solid) and the JK method (blue circle) in each tomographic bin. Although the JK method slightly overestimates the diagonal components for the second source bins at small scales, the overall amplitudes exhibit a good agreement with theory over the scales of interest.

\subsection{Boost factors}
\label{sec:measurement.Systematics.boost}

\begin{figure*}
\centering
\includegraphics[width=1.0\textwidth]{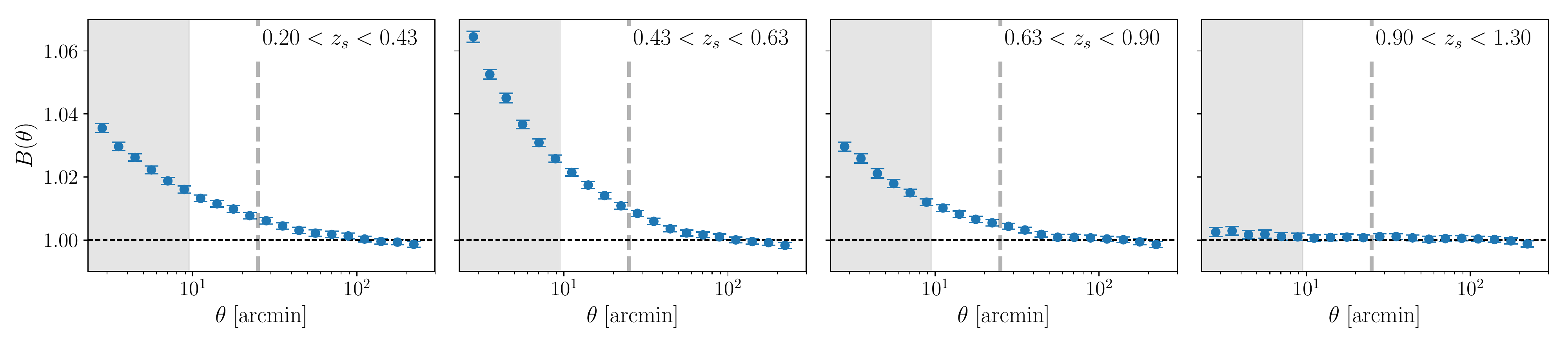}
\caption{ Boost factors estimated for each lens-source bin. The shaded region is the small scale that is removed by the $4 \hMpc$ scale cut. The vertical dashed lines indicate the scale cut of $12 \hMpc$. The second source bin shows the most significant impact of $\sim 7\%$ on the smallest scales.  
With the scale cut of $4\hMpc$, the level of the dilution reduces to below $3 \%$. It is below $1\%$ for the scale cut of $12\hMpc$.} 
\label{fig:boostfactor}
\end{figure*}

The mean tangential shear predicts lensing signals assuming galaxies are distributed on the sky homogeneously. However, since galaxies are clustered on small scales, 
sources behind the lenses could possibly be located closer to the lenses than predicted or physically associated with the lenses. 
These sources are less lensed than predicted or not lensed at all. Hence, they cause a dilution of the observed lensing signal \citep{Sheldon2004}. 
The extent of this contamination is estimated as the excess in the number counts of source galaxies in the region of lens galaxies compared to the random points distributed homogeneously. The excess for correcting this contamination (''boost factor'') is defined as 
\bea
B(\theta) = \frac{N_{\rm r} \sum_{\rm ls} w_{\rm ls}}{ N_l \sum_{\rm rs} w_{\rm rs}}~,
\eea
where $w_{\rm ls}~(w_{\rm rs})$ is the weight for the lens-source (random-source) pair, 
$N_{\rm l}~(N_{\rm r})$ is the total number of lenses (randoms).
Figure \ref{fig:boostfactor} shows the boost factors estimated for each source bin. 
The boost factor from the second source bin has the most significant impact of $\sim 7\%$ on the smallest scales 
due to the large fraction of galaxies overlapped in redshift distributions between lenses and sources, as shown in Figure \ref{fig:nz}. With the scale cut of $4\hMpc~(12\hMpc)$, the level of the dilution reduces to below $3 \%$ $(1 \%)$. Boost factors shown in this work are consistent with the results in \citet*{Prat2018DarkLensing}, computed from their third lens bin $( 0.45 < z < 0.60 )$ and the same source bins used in this work. 
The error bars are estimated by the JK calculation. We have corrected the measurements for the boost factors before the observing condition tests and the final analysis. 


\subsection{Potential Systematics}
\label{sec:measurement.Systematics}

In this section, we follow the procedures outlined in \citet*{Prat2018DarkLensing} to identify and correct for systematic biases correlating with galaxy-galaxy lensing. 
In Section \ref{sec:measurement.Systematics.null}, 
we compute the mean cross-component of the shear that should be consistent with zero if there are no potential systematics impacting our measurement. 
Potential uncertainties that may arise due to the redshifts of DMASS and the intrinsic alignments are explained in Section \ref{sec:measurement.Systematics.photoz} and Section \ref{sec:measurement.Systematics.ia}, respectively. In Section \ref{sec:measurement.Systematics.observing}, we investigate the impact of observing conditions. 

Since we utilize the same sources as \cite*{Prat2018DarkLensing}, we do not perform tests for systematics solely related to the shape estimation of sources. For source-specific tests, we refer readers to the tests of PSF residuals and Size \& S/N split described in \cite*{Prat2018DarkLensing}. The biases and uncertainties in photo-$z$ and the multiplicative shear for the same sources are discussed in  \cite*{Prat2018DarkLensing} and \cite{Troxel2018}.

\subsubsection{Cross component}
\label{sec:measurement.Systematics.null}

\begin{figure*}
\centering
\includegraphics[width=1.0\textwidth]{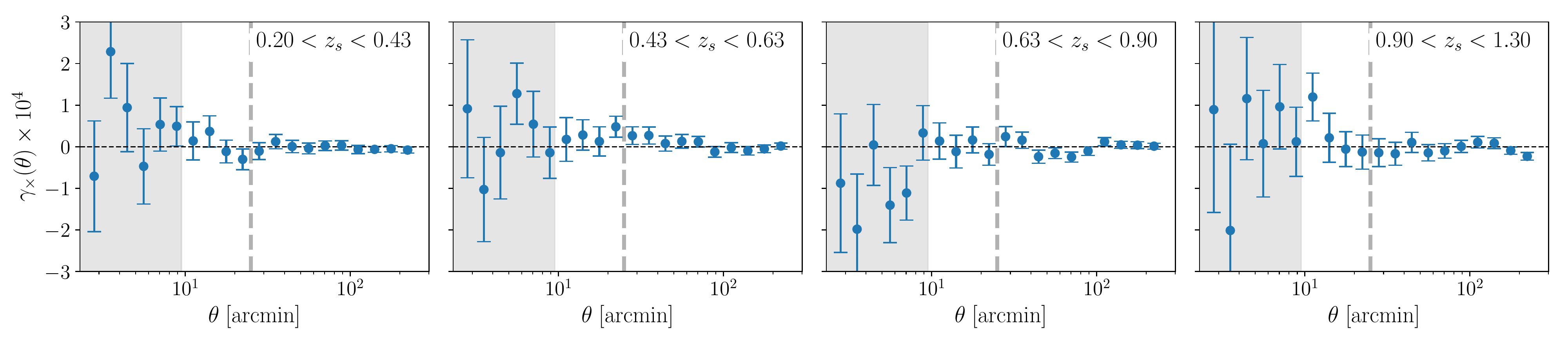}
\caption{ 
Mean cross-component of the shear for each lens-source bin pair. The shaded region is the small scale that is removed by the $4 \hMpc$ scale cut. The vertical dashed lines indicate the scale cut of $12 \hMpc$. The signals are consistent with zero above these scale cuts. 
   }
\label{fig:gammax}
\end{figure*}

The mean cross-component of the shear $\gamma_{\times}$ is a $45 \deg$ rotated signal with respect to the tangential shear $\gamma_{\rm t}$. If the shear is generated only due to the gravitational lensing, its cross-component should give a zero signal in the absence of systematic shear. The cross-component of shear is calculated using an equation equivalent to Equation \eqref{eq:gamma}.
%
The measured signal is subtracted by the signal around random points to remove additive contributions caused by geometrical effects.  


To quantify consistency with zero, we compute the $\chi^2$ of the null hypothesis given as 
\bea
\chi^2_{\rm null}  = \sum_{i,j} {\bf{d}}_i ({\rm {\bf{C}}}^{-1} )_{ij}{ \bf{d}}_j ~,
\label{eq:chi_null} 
\eea
where $d_i$ is the $i$th component of an observable to test, ${\rm C}$ is the corresponding covariance matrix. 
The result is shown in Figure \ref{fig:gammax}.  
We obtained $\chi_{\rm null}^2/{\rm dof} = 48.7/56$ for the scale cut of $4 \hMpc$ and $\chi_{\rm null}^2/{\rm dof} = 33.7/40$ for the scale cut of $12 \hMpc$.
As shown in Figure \ref{fig:gammax} and the values of $\chi^2_{\rm null}$, we have not detected any significant contributions of systematics from this test.

\subsubsection{ Redshift uncertainties in DMASS}
\label{sec:measurement.Systematics.photoz}

The redshift distribution of the DMASS sample is evaluated in \cite{DMASS} by the 'clustering-$z$' technique, which is the method that infers redshift distributions of an unknown sample by cross-correlating it with a galaxy sample whose redshift distribution is known and accurate. For further descriptions about the clustering-z method, we refer interested readers to \cite{Davis2017},  \cite{Cawthon2017DarkCross-Correlations}, \cite{Gatti2018}, and references therein. \cite{DMASS} utilizes the DES \redmagic sample \citep*{Rozo2016RedMaGiC:Data, ELVINPOOLE} as a reference sample. The \redmagic galaxies are red luminous galaxies selected by the redMaPPer algorithm \citep{Rykoff2014RedMaPPer.Catalog}, and have excellent photometric redshifts with an approximately Gaussian scatter of $\sigma_z/(1+z) < 0.02$. 
\cite{DMASS} finds a good agreement between the clustering-$z$ distribution of DMASS and the spectroscopic redshift of CMASS in the South Galactic Cap (SGC), as presented in Figure \ref{fig:nz}. 

The redshift distribution of a galaxy sample is modeled through the relation given as
\bea
n_{\rm true}(z) = \hat{n}(z - \Delta z)~,
\label{eq:photoz}
\eea
where $\hat{n}$ is the measured redshift distribution, and $\Delta z$ is the difference in the mean redshift of the true and measured distribution.
Utilizing the spectroscopic redshift distribution of CMASS as the true redshift distribution, \cite{DMASS} constrains 
$\Delta z$, the difference in the mean redshift of the CMASS and DMASS samples in this case, by jointly fitting the residuals of the angular correlation and clustering-$z$ measurements. The resulting number is $\Delta z = 3.5 \times 10^{-4}$ with its uncertainty of $\sigma_{\Delta z} = 0.5 \times 10^{-3}$. 
To incorporate the redshift uncertainty of DMASS in our analysis, we construct a Gaussian function whose standard deviation (std) is $\sigma_{\Delta z} $, and utilize the function as a prior for $\Delta z$. 

\subsubsection{Intrinsic alignments}
\label{sec:measurement.Systematics.ia}

The intrinsic alignment (IA) signal in galaxy-galaxy lensing is induced by contamination from source galaxies physically associated with the lens (for reviews, see \citepp{Troxel2015} and \citepp{Joachimi2015}). 
Red elliptical galaxies that form in primordial tidal fields tend to be radially aligned towards over-densities \citep{Hirata2007}. 
If galaxies physically associated with a lens are mistakenly assigned behind the lens due to significant redshift error, 
the alignments of those galaxies by the tidal field associated with the lens may introduce a negative signal, which reduces the measured galaxy-galaxy lensing signal. 

We parameterize the effects of IA using the nonlinear alignment (NLA) model \citep{Bridle2007}. This model impacts the lensing efficiency $W_{\kappa}$ as
\bea
W_{\kappa}^i(\chi) \to W_{\kappa}^i (\chi) - A ( z(\chi)) \frac{n_{\kappa} ( z(\chi))}{\bar{n}_{\kappa}} \frac{\ud z}{\ud \chi}~,
\eea
with 
\bea
A(z) = A_{\rm IA} \left(  \frac{1+z}{1+z_0} \right)^{\eta_{\rm IA}} \frac{0.0139 \Omega_m}{D(z)}~,
\eea
where $D(z)$ is the linear growth factor and $z_0=0.62$.  The amplitude of the intrinsic alignment $A_{\rm IA}$ and the scaling factor $\eta_{\rm IA}$ are treated as free parameters of the model. 


\subsubsection{Observing conditions}

\label{sec:measurement.Systematics.observing}
\begin{figure}
\centering
\includegraphics[width=0.45\textwidth]{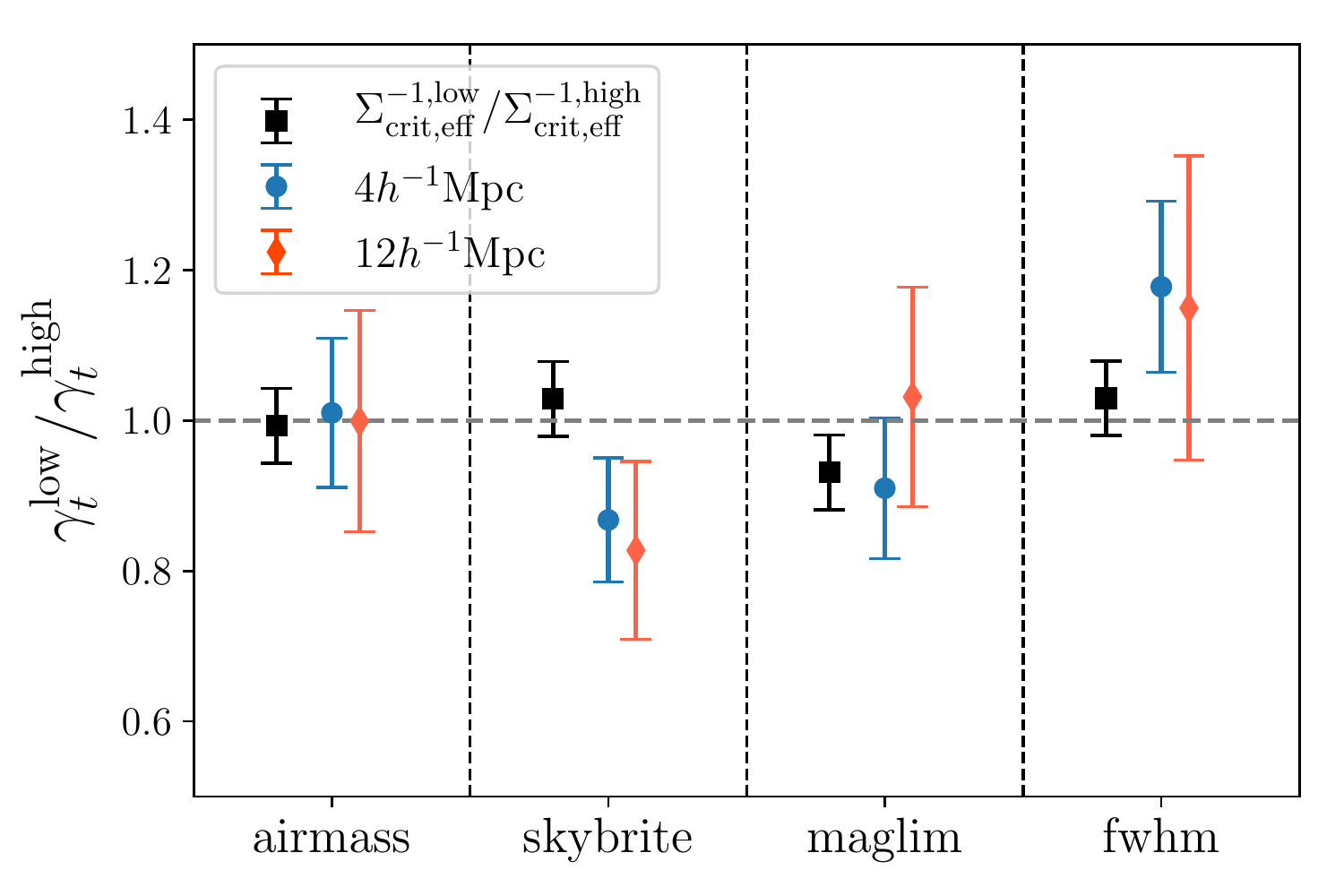}
\caption{The impact of observing conditions. Starting from the left, the properties listed on the x-axis are airmass, sky brightness (\texttt{skybrite}), $10\sigma$ limiting depth in the $r$ band \texttt{maglim}, and seeing FWHM. The black square points show the ratio of $\Sigma_{\rm crit, eff}^{-1}$ using the redshift distributions of sources in each split region. The blue (red) points are the ratio between the amplitudes fitted with the theoretical tangential shear prediction for each half, with the scale cut of $4 \hMpc ~(12 \hMpc)$.  
}
\label{fig:observing}
\end{figure}

In this section, we examine potential biases in galaxy-galaxy lensing that may arise due to observing conditions. 
The impact of observing conditions on lenses only is discussed in Section 4 in \cite{DMASS}. 
The resulting set of weights from that analysis has been applied to lenses before further testing. 

To search for potential systematic uncertainties associated with observing conditions, we follow the methodology described in \citet*{Prat2018DarkLensing}.
We use HEALPix maps ($N_{\rm side} = 4096$) of airmass, seeing FWHM, sky brightness (\verb+skybrite+) and $10\sigma$ limiting depth (\verb+maglim+) in the $r$ band. A detailed description of constructing HEALPix survey property maps can be found in \cite{Leistedt2015}. 
Using each HEALPix map, we split the source and lens galaxy samples  into halves of either low or high values of a given quantity.
Then, we compute the galaxy-galaxy lensing signal in each half, and examine the discrepancies between two signals. The sources are all combined into a single bin to maximize the sensitivity to potential differences between the halves. 


The observing conditions might be weakly correlated with photo-$z$ of lenses or sources. The correlations with photo-$z$ would result in a difference in the mean redshift of the split source samples and thereby affect the amplitude of the lensing signals. 
For \texttt{maglim}, the correlation with photo-$z$ results in a difference of 0.03 in the mean redshift. For other observing conditions, the differences are of the order of $0.01$ or smaller. 
However, removing the correlations with photo-$z$ should be treated in the catalog level and is beyond the scope of this paper. Therefore, we do not correct contamination related to photo-$z$, but instead estimate what contribution it has to any apparent systematic bias.
To separate the impact of systematics that we want to correct from the photo-$z$ related systematics, we utilize the geometric factor $\Sigma_{\rm crit}^{-1}$. The geometric factor takes into account the difference in the amplitude of the lensing signals  due to the redshift distributions and thereby enables us to predict the impact of photo-$z$ related systematics. We compute the ratio of the geometric factor $\Sigma_{\rm crit}^{-1}$ with the source redshift distribution of each of the halves, and compare the quantity with the ratio of the tangential shear signals. 
For the lenses, we simply use the same redshift distribution for the two halves as the difference in the mean redshift of the lenses is negligible. 

The geometric factor $\Sigma_{\rm crit}^{-1}$ is defined as
\bea
\Sigma_{\rm crit}^{-1}(z_{\rm l}, z_{\rm s}) = \frac{4\pi G}{c^2} \frac{ D (z_{\rm l}, z_{\rm s}) D (z_{\rm l})}{ D (z_{\rm s}) }~,
\eea
where  $D(z)$ is the angular diameter distance to the redshift $z$, $z_{\rm l}$ and $z_{\rm s}$ are the redshifts of lens and source galaxies. The geometric factor becomes zero for $z_{\rm s} < z_{\rm l}$.  
The width and overlap of the redshift distributions can be incorporated by integrating the geometric factor over the redshift range of lens and source bins as 
\bea
\Sigma_{\rm crit,eff}^{-1}(z_{\rm l}, z_{\rm s}) = \int \int \ud z_{\rm l} \ud z_{\rm s} n_{\rm l} (z_{\rm l}) n_{\rm s} (z_{\rm s}) \Sigma_{\rm crit}^{-1}(z_{\rm l}, z_{\rm s})~.
\eea
The effective geometric factor $\Sigma_{\rm crit,eff}^{-1}$ can be related to the tangential shear as
\bea
\gamma_{\rm t} = \frac{\Delta \Sigma}{\Sigma_{\rm crit, eff}}~, 
\eea
where $\Delta \Sigma$ is the excess surface mass density. 
If the measured signal is independent of a survey property, we expect the ratio of the effective geometric factor to be the same as the ratio of the tangential shear:
\bea
\frac{  \Sigma_{\rm crit,eff}^{-1, {\rm high} } }{  \Sigma_{\rm crit,eff}^{-1, {\rm low} } }
=
\frac{\gamma_{\rm t}^{\rm high} }{  \gamma_{\rm t}^{\rm low} }~.
\eea 
Note that the geometric factor ratio is reduced to unity if the survey property is not correlated with photometric redshift.
To minimize possible biases arising while fitting two noisy quantities, we fit an amplitude of each signal to the theoretical prediction using the scales chosen and then compute a ratio of these fitted amplitudes.

\begin{table*}
\centering
\caption{Parameters and priors used to describe the measured galaxy-galaxy lensing signal. `Flat' is a flat prior in the range given while `Gauss' is a Gaussian prior with mean $\mu$ and width $\sigma$. Priors for the tomographic shear and photo-$z$ bias parameters $m^i$ and $\Delta z_{\rm src}^i$ are identical to the DES Y1 analysis \citep{DESCollaboration2017}.}
\label{tab:params}
\begin{tabular}{@{}llllllclllclllcll@{}}
\toprule
 &  & \multicolumn{1}{c}{Parameter} &  &  &  & Notation                      &  &  &  & Fiducial &  &  &  & Prior                 &  &  \\ \midrule
 &  & Galaxy bias (galaxy clustering)      &  &  &  & $b_{\rm g}$                          &  &  &  & 2.0      &  &  &  & Flat (0.8, 3.0)       &  &  \\
 &  & Galaxy bias (galaxy-galaxy lensing)             &  &  &  & $b_{\gamma}$                   &  &  &  & 2.0      &  &  &  & Flat (0.8, 3.0)       &  &  \\
 &  & Correlation coefficient       &  &  &  & $r_{\rm cc} ~(=b_{\gamma}/b_{\rm g})$ &  &  &  & 1.0      &  &   &  &           $\cdot$           &  &  \\
 &  & Intrinsic alignment amplitude          &  &  &  & $A_{\rm IA}$                   &  &  &  & 0.0      &  &  &  & Flat (-5.0, 5.0)      &  &  \\
 &  & Intrinsic alignment scaling          &  &  &  & $\eta_{\rm IA}$                &  &  &  & 0.0      &  &  &  & Flat (-5.0, 5.0)      &  &  \\
 &  & Lens redshift bias            &  &  &  & $\Delta z_{\rm lens}$          &  &  &  & 0.0035      &  &  &  & Gauss ( 0.0035, 0.005)   &  &  \\
 &  & Source photo-$z$ bias ($i=1$)     &  &  &  & $\Delta z^1_{\rm src}$         &  &  &  & -0.001   &  &  &  & Gauss (-0.001, 0.016) &  &  \\
 &  & Source photo-$z$ bias ($i=2$)     &  &  &  & $\Delta z^2_{\rm src}$         &  &  &  & -0.009   &  &  &  & Gauss (-0.009, 0.013) &  &  \\
 &  & Source photo-$z$ bias ($i=3$)     &  &  &  & $\Delta z^3_{\rm src}$         &  &  &  & 0.009    &  &  &  & Gauss (0.009, 0.011)  &  &  \\
 &  & Source photo-$z$ bias ($i=4$)     &  &  &  & $\Delta z^4_{\rm src}$         &  &  &  & -0.018   &  &  &  & Gauss (-0.018, 0.022) &  &  \\
 &  & Shear calibration bias $(i\in\{1,2,3,4\})$      &  &  &  & $m^i$                          &  &  &  & 0.012    &  &  &  & Gauss (0.012, 0.023)  &  &  \\ \bottomrule
\end{tabular}
\end{table*}

The results are displayed in Figure \ref{fig:observing}. The black square points with error bars show the ratio of $\Sigma_{\rm crit, eff}^{-1}$ using the redshift distribution of sources in each split region. 
The black point of the \texttt{maglim} case shows a slight deviation from unity which implies that photo-$z$ of sources is weakly correlated with the observing condition as expected from the difference in the mean redshift of the split source samples. 
The blue (red) points with error bars are the ratio between the amplitudes fitted with the theoretical prediction of tangential shear for each half with the scale cut of $4\hMpc~(12\hMpc)$. The size of the error bars is computed by the JK method. 
The blue point of the \texttt{maglim} case shows the same 
 deviation from unity as the black point, which indicates that the correlation with photo-$z$ is the main source of systematics related to \texttt{maglim}. 
However, both the blue and red error bars for the same case are consistent with the line of unity and the black error bar simultaneously, which implies that this photo-$z$ related systematics is well below the statistical uncertainty.
For the case of \texttt{skybrite}, the blue and red points show a mild difference of $1$--$2\sigma$ from the black point, which is not statistically significant enough to warrant further action. For the rest of the properties, the blue/red points and black point show a good agreement.
Hence, we conclude that we do not observe any significant impact of observing conditions and thereby do not correct them.

\subsection{  Likelihood Analysis }
\label{sec:measurement.likelihood}

Using a combination of galaxy-galaxy lensing and galaxy clustering, we perform Markov Chain Monte-Carlo likelihood analyses to constrain the parameter set of \{$b_{\rm g}$, $b_{\gamma}$\} in fixed cosmology. The cross-correlation coefficient $r_{\rm cc}$ is derived from the ratio of the two galaxy bias constraints. 
Along with the parameter set, we also vary nuisance parameters describing the shear and photo-$z$ systematics for different tomographic bins, and model parameters for the intrinsic alignment. 
Since we use an identical source sample as the DES Y1 analysis \citep{DESCollaboration2017}, we adopt the same models for the shear and photo-$z$ systematics. 
The complete set of varied parameters and priors is summarized in Table \ref{tab:params}. 

The likelihood of the combined probe is evaluated by the sum of individual log-likelihoods given as  
\bea
\ln \mathcal{L}(p) = - \frac{1}{2} \left[ \chi^2_{\rm g\kappa}(p) + \chi^2_{\rm gg}(p)\right]~,
\eea
where $p$ is the set of varied parameters, the subscript `gg' represents galaxy clustering of BOSS CMASS,  and `g$\kappa$' denotes galaxy-galaxy lensing of DMASS. 
We assume there is no cross-correlation between two probes as the two survey areas do not overlap\footnote{Sources in the overlapping area between DES and BOSS were used to train the DMASS algorithm. Afterwards, those sources were excluded from the final DMASS sample.}. 
We estimate the value of $\chi^2$ as below:
\bea
\chi^2 = \sum_{i,j} ( {\bf{d}} - {\bf{d}}_{\rm th}  ) _i {\rm {\bf{C}}}^{-1}_{ij} ( {\bf{d}} - { \bf{d}}_{\rm th}  )_j^T ~,
\eea
where $d_{\rm th}$ and $d$ are theoretical and measured datavector respectively. To compute the value of $\chi^2_{\rm g\kappa}$, Equation \eqref{eq:gammat} is adopted as a theoretical datavector, and its corresponding covariance matrix is described in Section \ref{sec:measurement.CovarianceMatrix}. For galaxy clustering, we use a set of values of $\{ H(z), d_{\rm A}(z), \Omega_{\rm m} h^2, f(z)\sigma_8(z), b\sigma_8(z) \}$ at redshift $z=0.59$ as a datavector with correlations between those observables described in Section \ref{sec:data.cmass}. 

To evaluate the likelihood values and matter power spectrum for a given cosmology, we use the DES analysis pipeline in CosmoSIS \citep{COSMOSIS}. Further details of the likelihood framework are illustrated in \cite{Krause2017}.

\subsection{Blinding}
\label{sec:measurement.blinding}

We blinded the results to protect against human bias. The cosmological parameter constraints were plotted with shifted axes. No comparison to theory predictions at the two-point level ($\gamma_{\rm t}$) or of cosmological contours was made. 
In order to interpret the results objectively while avoiding  confirmation bias, we prepared two different versions of the result section for two possible scenarios $-$ the case where $r_{\rm cc}$ is consistent with unity within $1\sigma$ and the opposite $-$ before unblinding, so we can choose which version of the results to use depending on the unblinded result. 
We unblinded after we ensured that there are no major systematics that can bias the cosmological constraints through various tests listed in Section \ref{sec:measurement.Systematics}. 
No change was made in either the analysis method or pipeline after unblinding.

%% file: result_ggl.tex
\begin{figure}
\centering
\includegraphics[width=0.48\textwidth]{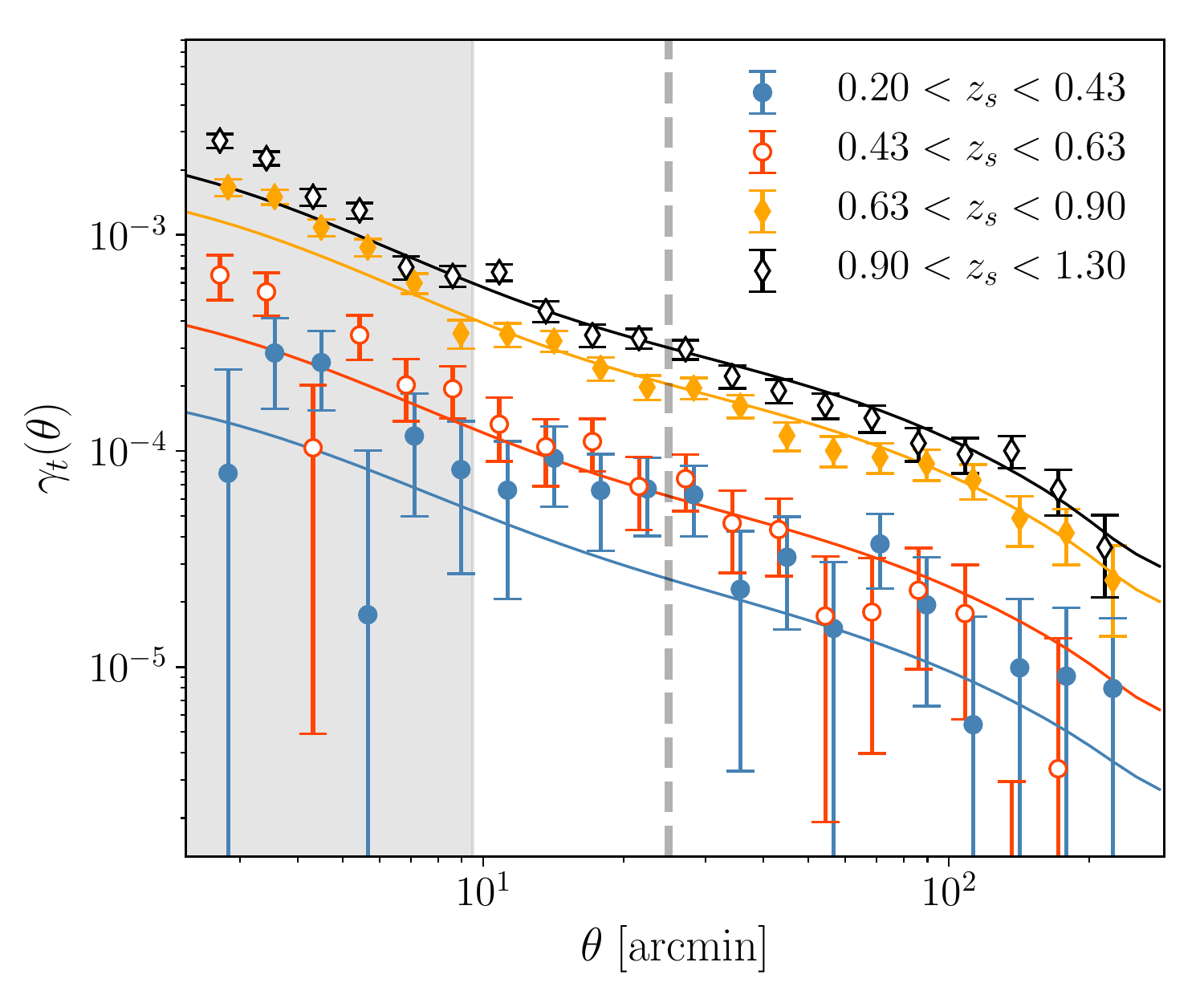}
\caption{
Tangential shear signals measured with the DMASS lenses and \metacal\ sources. The solid lines are the best-fit theory lines. The shaded region is the scales removed by the $4 \hMpc$ scale cut. The vertical dashed line indicates the scale cut of $12 \hMpc$. 
}
\label{fig:gammat}
\end{figure}

\section{Results}
\label{sec:result}

In this section, we present the details of our measurement of galaxy-galaxy lensing and the cross-correlation coefficient $r_{\rm cc}$ with the discussion about the implication of the results. 

\subsection{Tangential shear measurements}
\label{sec:result-gammat}

In Figure \ref{fig:gammat}, we present the measurement of tangential shear from DMASS and \metacal\ in four different tomographic bins (colored points with error bars). 
Solid lines are theoretical predictions from our fiducial cosmology with the best-fitting galaxy bias $b_{\rm g}$ and correlation coefficient $r_{\rm cc}$ (the values of these parameters are presented in Table \ref{tab:bias}).
Statistical errors are obtained from the theoretical covariance matrices estimated in Section \ref{sec:measurement.CovarianceMatrix}. 
The values of $\chi^2/$dof against the theoretical predictions are calculated as $\chi^2/ {\rm dof} = 49.6/56$ for the scale cut of $4 \hMpc$, and $\chi^2/ {\rm dof} = 36.2/40$ for the scale cut of $12 \hMpc$. 
The signal-to-noise ratio of the measured tangential shear is calculated using the equation $S/N= ( {\bf{d}} { \bf{C}} ^{-1} { \bf{d}}^T) ^{1/2}$, 
where $\bf{d}$ is the vector of $\gamma_{\rm t}$ in each angular bin and $\bf{C}$ the covariance matrix. Our overall lensing signal is detected with $S/N = 25.7$ using the scales $r > 4 \hMpc$, and $S/N =17.7$ for $r > 12 \hMpc$. 

As shown in the figure, the measured tangential shear with the lowest source bin $(0.2 < z < 0.43)$ is slightly higher than the best-fit theory obtained by fitting all of the four tangential shear signals simultaneously with one galaxy bias parameter. This indicates that the lowest tangential shear favors a higher galaxy bias $b_{\gamma}$ than other signals and CMASS. We compute $\chi^2$ of the lowest tangential shear alone varying galaxy bias and find that the value of galaxy bias that minimizes $\chi^2$ is $b_{\gamma} = 4.0$, which is nearly twice as high as that of CMASS.
A potential reason for this can be found in the original CMASS sample. \cite{Salazar-Albornoz2017} measured the galaxy bias of CMASS in fine redshift bins and found that galaxy bias peaks at the low-redshift end ($z \sim 0.45$) instead of increasing monotonically (see Figure 7 in their paper). As the DMASS algorithm works relatively poorly near the edge of low redshifts \citep{DMASS}, it is possible that the irregularity of galaxy bias at low redshifts might be amplified while the algorithm faithfully replicates the properties of CMASS. If the same irregularity exists in the DMASS sample, the impact can be shown significantly in the tangential shear signal from the lowest source bin because the lowest source bin of DES Y1 is located in front of the DMASS lens bin as shown in Figure \ref{fig:nz}. In this case, the signal only captures correlations with the DMASS sample at low redshifts where the two samples partially overlap. The constraining power on the galaxy bias is mainly coming from the higher redshift source bins which is weighting the high-redshift end of the full DMASS sample more. 
Therefore, we do not correct the galaxy bias model in this work.
There will be further discussion about the impact of the lowest tangential shear on the constraint on galaxy bias in the next section. Future high precision analyses will likely need to model this behavior of the galaxy bias when using the DMASS sample as lenses. 

%
The residual systematics in the source redshifts could possibly contribute to the mismatch. 
The redshift distributions for DES have been obtained by using data from the 30-band photometric data set `COSMOS-2015' (Laigle et al. 2016). 
%
However, \cite{Joudaki2020} and \cite{Hildebrandt2020} have found a coherent downward shift in the redshift distributions between COSMOS-2015 and spectra due to the `catastrophic outlier' fraction of $~6\%$ in the magnitude range $23 < i < 24$ reported in \cite{Laigle2016}. 
\cite{Alarcon2021} have also found photo-$z$s of COSMOS-2015 to be biased towards lower redshifts with respect to the spectroscopic sample, with a larger bias at higher redshift and fainter magnitudes. 
This could impact the DES Y1 source redshift distributions, especially for the shape of the high-$z$ tail  where the lens and the first source bin overlap.

\subsection{Cross-correlation coefficient $r_{\rm cc}$}
\label{sec:result-rcc}

In this section, we present the measurements of  $r_{\rm cc}$ from 
jointly fitting galaxy clustering and galaxy-galaxy lensing using the MCMC fitting method. We use the BOSS CMASS galaxies for galaxy clustering and the DMASS galaxies for galaxy-galaxy lensing. Note that we perform this analysis in fixed cosmology because the primary motivation for this paper is to quantify the difference in galaxy bias from the two probes, not to constrain the galaxy bias itself.

Fixing the cosmology to that of  {\it Planck} 2018, 
we first constrain the galaxy clustering bias $b_{\rm g}$ from BOSS CMASS galaxy clustering to detect any potential biases that may appear due to our fiducial pipeline. We obtain $b_{\rm g} \sigma_8(z=0.59) = 1.154 \pm 0.080$ with a fixed value of $\sigma_8(z=0.59)=0.60$. 
This value is consistent with $b_{\rm g}\sigma_8 (z=0.59) = 1.154 \pm 0.090$ from the published BOSS measurement \citep{Chuang2017}. This also shows that the analysis of this work is not sensitive to our choice of fiducial cosmology. 
From galaxy-galaxy lensing alone, we obtain the galaxy lensing bias $b_{\gamma} = 2.04^{+0.16}_{-0.16}$ for the fiducial scale cut of $12 \hMpc$, and $b_{\gamma} =2.10^{+0.13}_{-0.12}$ for the scale cut of $4 \hMpc$. 

\begin{figure}
\centering
\includegraphics[width=0.45\textwidth]{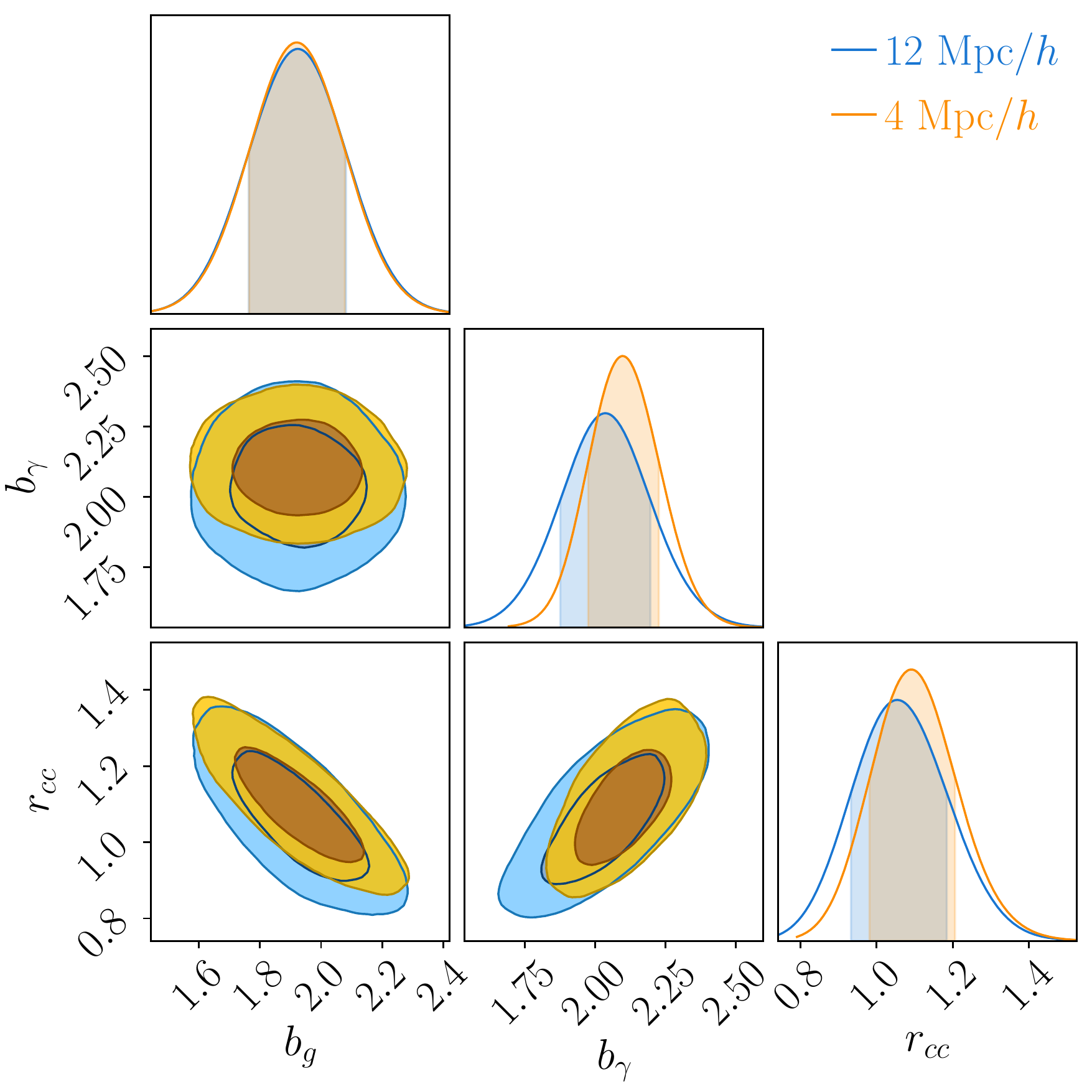}
\caption{ 
Constraints on galaxy bias from the BOSS galaxy clustering signal ($b_{\rm g}$), from the DMASS galaxy-galaxy lensing signal ($b_{\gamma}$), and the correlation coefficient ($r_{\rm cc}$) derived from the ratio of the two galaxy biases. 
We find that the galaxy bias inferred from the DMASS galaxy-galaxy lensing signal is consistent with the galaxy bias of BOSS CMASS. The derived value of $r_{\rm cc}$ is consistent with unity for both scale cuts. 
 }
\label{fig:b_ggl}
\end{figure}

Next, the cross-correlation coefficient $r_{\rm cc}$ is measured by jointly 
fitting the galaxy-galaxy lensing measurement of DMASS 
with the results of galaxy clustering in BOSS CMASS, parametrized as $\{H(z), d_{\rm A}(z), \Omega_{\rm m} h^2, f(z)\sigma_8(z), b\sigma_8(z) \}$ at $z=0.59$. 
Figure \ref{fig:b_ggl} shows contours in a two dimensional plane of $b_{\rm g}$ and $r_{\rm cc}$ constrained using two different scale cuts. The blue contours show when the fiducial scale cut of $12\hMpc$ is applied. The orange contours are for the scale cut of $4 \hMpc$. 
We find that $r_{\rm cc} = 1.06^{+0.13}_{-0.12}$ and $b_{\rm g} = 1.92^{+0.16}_{-0.16}$ for the scale cut of $12 \hMpc$, and $r_{\rm cc} = 1.09^{+0.12}_{-0.11}$ and $b_{\rm g} = 1.92^{+0.16}_{-0.15}$ for the scale cut of $4 \hMpc$. All of these numbers are listed in Table \ref{tab:bias} as well. 
%
The constraints of $r_{\rm cc}$ favor a value slightly higher than unity for both scale cuts. 
These results indicate that $b_{\gamma}$ from DMASS is slightly higher than $b_{\rm g}$ from CMASS. However, they are consistent with unity within $1\sigma$, which implies that the discrepancy between the galaxy bias constraints of DMASS and CMASS and the effects of non-linearity/stochasticity in DMASS are well below the statistical uncertainties of the survey, over the scales $ > 4 \hMpc$. 
%
%
%
The mild preference of $r_{\rm cc}$ for a higher value shown in this work may be relieved with the DES Year 3 shape calibration. In DES Year 3 \citep{MacCrann2021}, the shear calibration bias prior is shifted from $m^i=0.012$ to $m^i = \{-0.0063, -0.0198, -0.0241, -0.0369\}$, where the subscript $i$ indicates $i$th source bin. The shift in the negative direction would result in increasing the amplitude of the tangential shear. Then, the galaxy bias is pulled down to compensate for the increase, which leads to a decrease in $r_{\rm cc}$.

\begin{figure}
\centering
\includegraphics[align=t,width=0.22\textwidth]{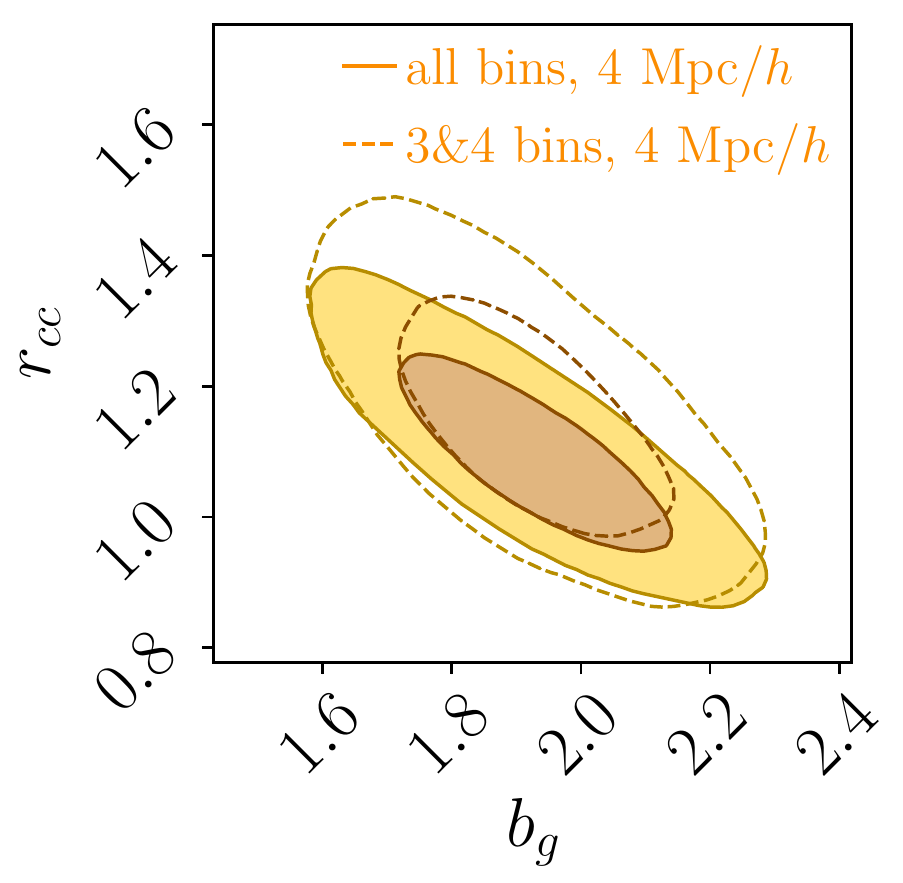}
\includegraphics[align=t,width=0.22\textwidth]{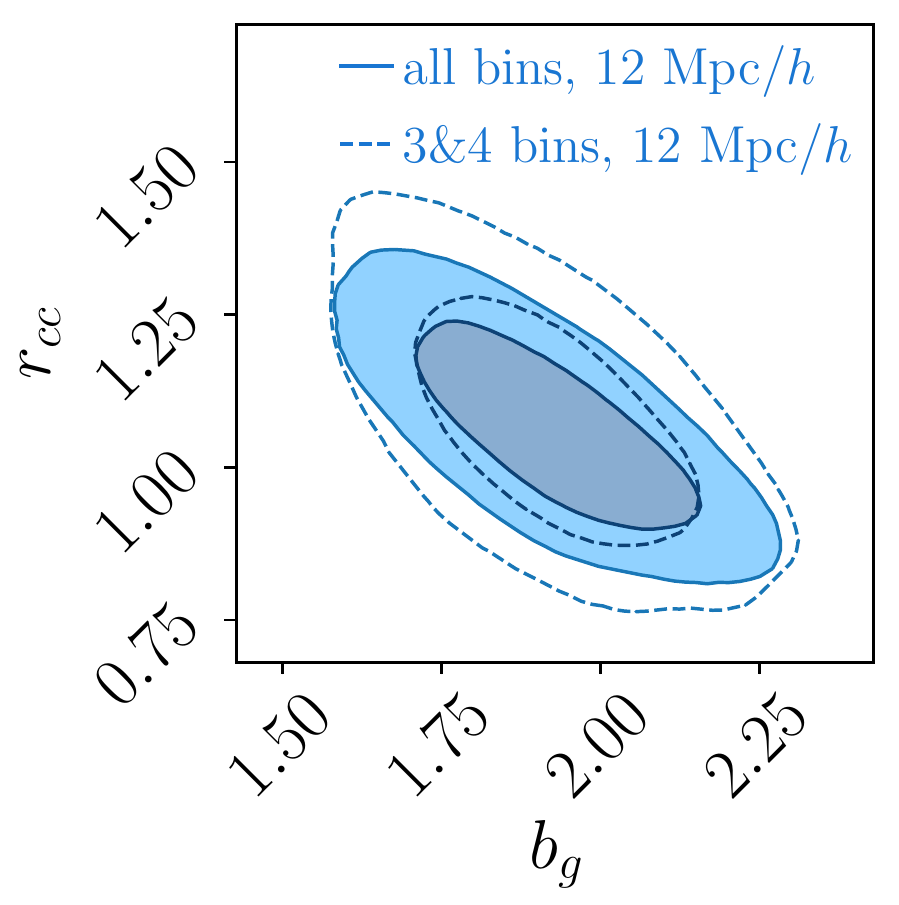}
\caption{ Galaxy bias and the correlation coefficient $r_{\rm cc}$ obtained using all tangential shear signals (solid) and only the signals meaured with the third $\&$ fourth bins (dashed). 
}
\label{fig:b_ggl_robustness2}
\end{figure}

We additionally test the robustness of our results. 
As the tangential shear signals measured with the first ($0.20 < z < 0.43$) and the second ($0.43 < z < 0.63$) source bins show a significantly low signal-to-noise ratio compared to the others (see Section \ref{sec:result-gammat}), 
we measure the constraints of $b$ and $r_{\rm cc}$ without the first two bins and compare them with the constraints obtained with all source bins. The results are presented in Figure \ref{fig:b_ggl_robustness2}. Each panel shows the constraint with all bins (solid lines) and without the first two bins (dashed lines) for different scale cuts. The resulting numbers are  $r_{\rm cc} = 1.15^{+0.14}_{-0.14}$ and $b_{\rm g} = 1.91^{+0.16}_{-0.15}$ for the scale cut of $4\hMpc$, and $r_{\rm cc} = 1.06^{+0.16}_{-0.15}$ and $b_{\rm g} = 1.92^{+0.17}_{-0.16}$ for the scale cut of $12\hMpc$. The constraints are slightly shifted towards higher values but still consistent within $1\sigma$. 
As stated in Section \ref{sec:result-gammat}, the tangential shear signal measured with the first source bin ($0.20 < z < 0.43$) is higher than predicted in theory due to the interplay between the first source bin being ahead of the lens bin and the irregularity of galaxy bias at low redshifts. This additional analysis also proves that the impact from the galaxy bias at low redshifts is negligible.

\begin{figure*}
\centering
\includegraphics[width=\textwidth]{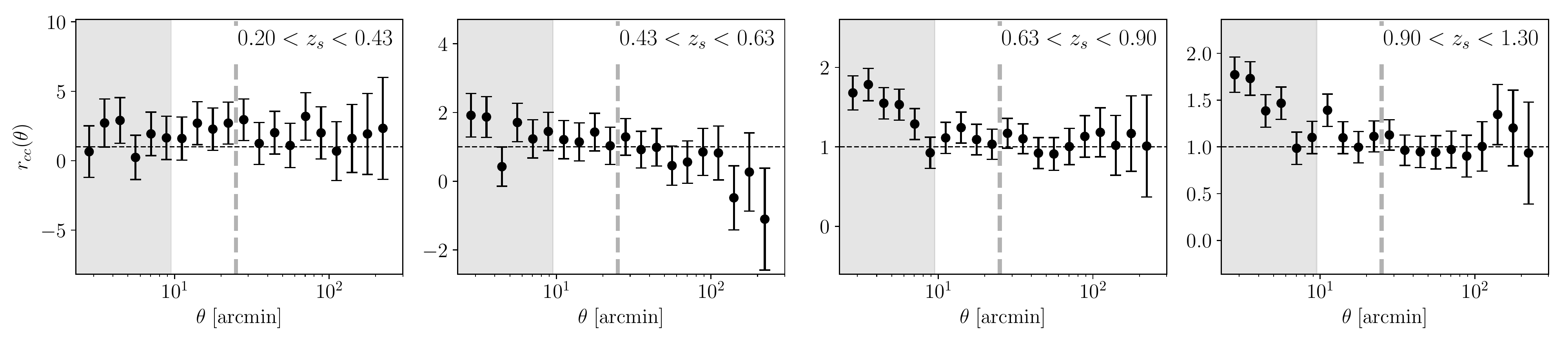}
\caption{ 
Cross-correlation coefficient as a function of angular separation obtained by dividing the measured tangential shear by theoretical predictions modeled with \texttt{halofit} and linear galaxy bias.  The dashed line is 1.0. The shaded region is removed by the $4 \hMpc$ scale cut. The dashed vertical line in grey denotes the scale cut of $12\hMpc$. The measured $r_{\rm cc}$ is consistent with unity for all scales for the first two bins and on the scales above $4\hMpc$ for the the last two. 
}
\label{fig:rcc_theta}
\end{figure*}

Finally, we evaluate the scale-dependence of the cross-correlation coefficient as a function of angular separations. 
Figure \ref{fig:rcc_theta} displays $r_{\rm cc}$ for different tomographic bins.  
The quantities are computed by dividing the measured tangential shear by theoretical predictions modeled with \texttt{halofit} \citep{Takahashi2012} implemented in \cosmosis\ and linear galaxy bias $b_{\rm g} = 2$. The dashed line shows the ideal case, unity. 
The shaded region is the small scale that is removed by the $4 \hMpc$ scale cut. 
The measured $r_{\rm cc}$ is consistent with the line of unity for all scales for the lowest two bins. 
For the highest two bins, we see a small discrepancy at small scales as expected, but overall the results show a good agreement with the line of unity above $4\hMpc$. 
The values of $\chi^2/$dof against unity are calculated as $55.9/56$ above  $4\hMpc$ and $37.7/40$ above $12\hMpc$.  


\subsection{Adding angular galaxy clustering}
\label{sec:wtheta}

\begin{figure}
\centering
\includegraphics[width=0.4\textwidth]{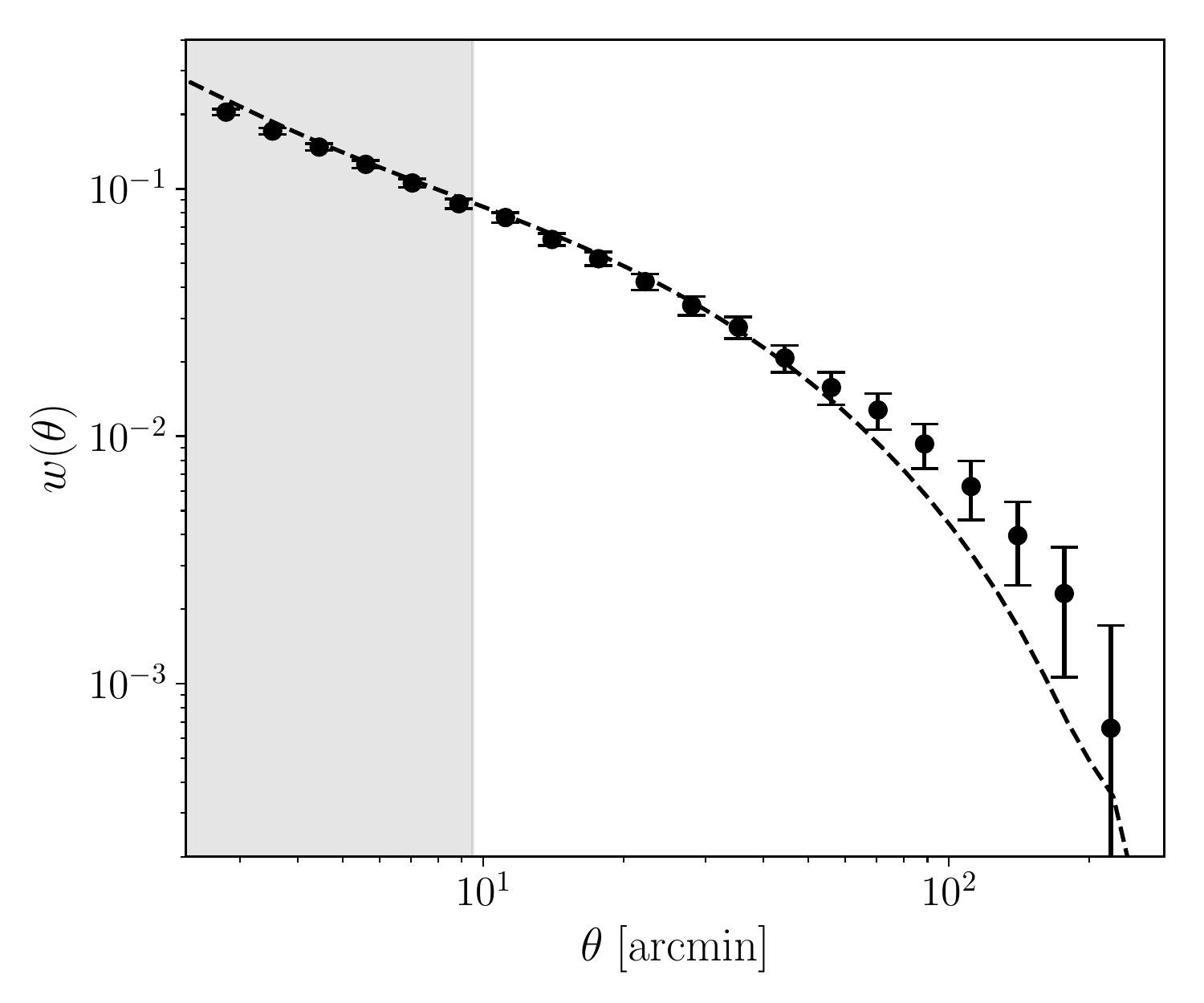}
\caption{ 
The angular galaxy clustering measurement of DMASS. The dashed line is the best-fitting prediction. The shaded region ($r < 4 \hMpc$) is discarded in the analysis to exclude the small scales where the nonlinear effect is significant. 
}
\label{fig:wtheta}
\end{figure}

\begin{figure}
\centering
\includegraphics[width=0.45\textwidth]{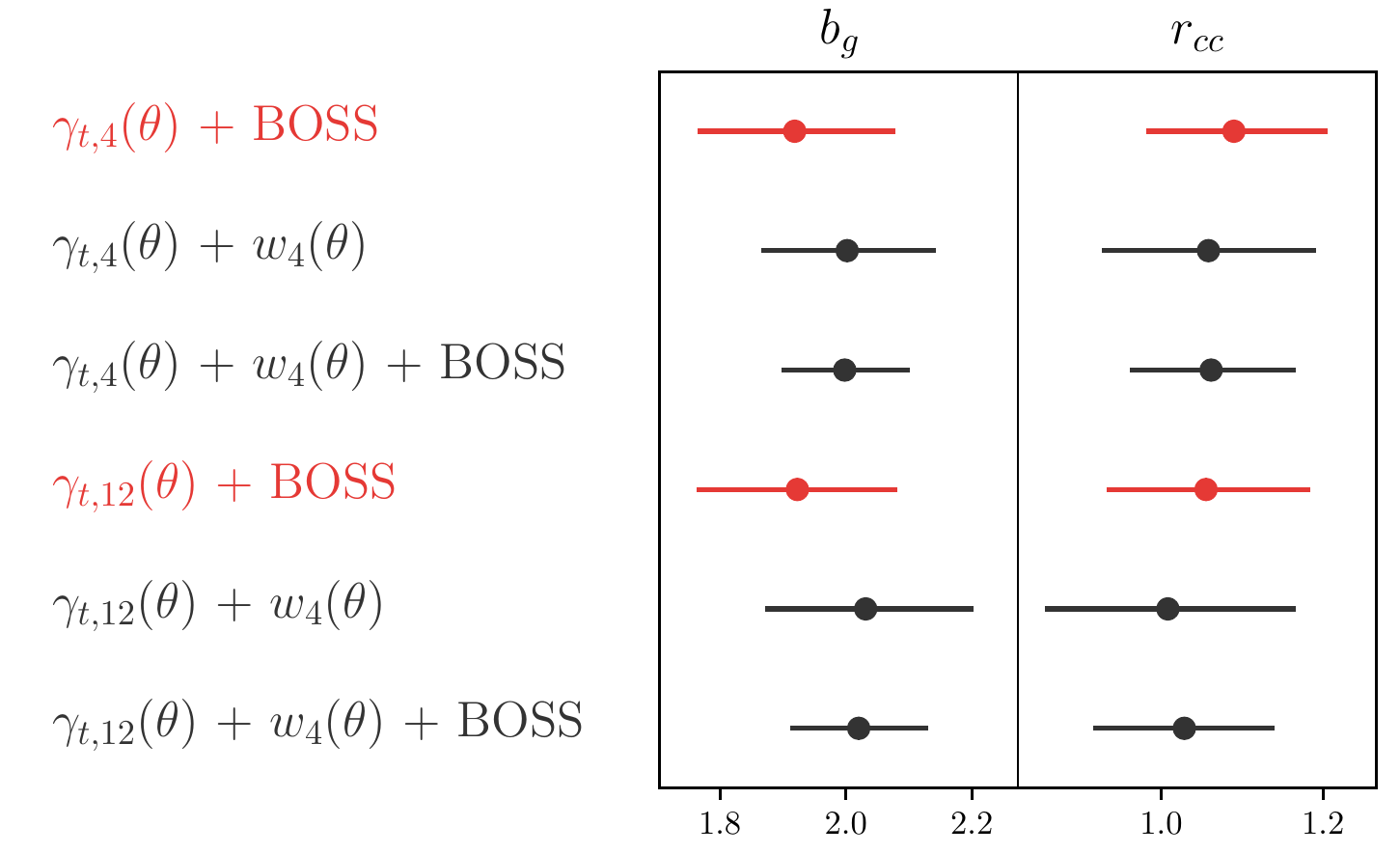}
\caption{ 
The constraints of $b_{\rm g}$ and $r_{\rm cc}$ obtained from the various combinations of data sets. The red error bars are obtained from the galaxy-galaxy lensing of DMASS combined with the BOSS CMASS data. The black error bars are obtained from the  angular clustering combined with other probes. The subscript `4' and `12' denote the cut-off scales $4 \hMpc$ and $12 \hMpc$, respectively. 
 }
\label{fig:errorbar_wtheta}
\end{figure}

In the previous section, we have restricted the number of data sets to be as minimal as possible to carefully examine the galaxy-galaxy lensing of DMASS without introducing potential systematic biases from other probes. Angular clustering is a powerful probe to constrain galaxy bias, but adding angular clustering of DMASS may dilute any potential issues coming from galaxy-galaxy lensing, and also require additional validations for the scale cut or covariances.   
%
However, it would be interesting to see the full statistical power from DES, assuming the simplest case. 
Hence, in this section, we present the constraint on $r_{\rm cc}$ measured with the  angular galaxy clustering of DMASS. 
%
The galaxy bias inferred from the angular galaxy clustering of DMASS is consistent with the galaxy bias of BOSS CMASS within $1\sigma$ \citep{DMASS}. Therefore, the angular galaxy clustering of DMASS will play the same role as the BOSS measurements but will convey the  constraining power from DES. 


The theoretical prediction for angular clustering is given as 
\bea
w(\theta) = \frac{1}{2\pi} \int^{\infty}_{0} C_{\rm gg} (\ell) J_{0} (\ell\theta) \ell \ud \ell  ~,
\label{eq:wtheta}
\eea
with the angular galaxy power spectrum  
\bea
C_{\rm gg} =  b_{\rm g}^2  \int^{\infty}_{0} \frac{\ud \chi}{\chi^2}  \left( \frac{n_{\rm g} (z(\chi)) }{\bar{n}_{\rm g}} \frac{\ud z}{\ud \chi} \right)^2 P_{\delta\delta} (k, z(\chi))~.
\label{eq:cgk}
\eea 
As shown in the above equation, the amplitude of angular galaxy clustering is proportional to $b_{\rm g}^2$, thereby adding angular clustering helps to break the degeneracy between $b_{\rm g}$ and $r_{\rm cc}$.

\cite{DMASS} measured the angular galaxy clustering of DMASS to validate that DMASS matches the BOSS CMASS sample. We recompute the signal using the exact same methodology but with the number of angular bins increased from 10 to 20. 
As we obtained the same results except for the number of bins, we only briefly summarize the methodology here and refer readers to the original paper. 
%
%
The correlation function was measured in 20 logarithmically spaced angular bins over the range $2.5 \arcmin < \theta < 250 \arcmin$. Weights for mitigating potential systematics are applied to each galaxy, which is illustrated in section 4 in \cite{DMASS}. 
The covariance matrix of angular clustering and galaxy-galaxy lensing is computed by \cosmolike\ as described in section \ref{sec:measurement.CovarianceMatrix}, including the cross-covariance between the two probes.
We assume there is no cross-correlation with BOSS as the two survey areas do not overlap. 
Galaxy clustering is less sensitive to the nonlinear effects at small scales than galaxy-galaxy lensing. Hence, we choose a more aggressive scale cut of $4 \hMpc$. This is a reasonable choice as \cite{DMASS} shows that the galaxy bias of DMASS is consistent with that of CMASS using the angular clustering of DMASS over the scales $> 2 \hMpc$. The measured signal is plotted with the best-fitting prediction in Figure \ref{fig:wtheta}. 
The small excess at large scales in the measurement is due to the RSD effect that is not included in the theoretical prediction. We find that CMASS angular clustering also shows a similar level of deviation from the best-fitting theory at the same scales. Despite the deviation, we obtain a reasonable value $\chi^2/{\rm dof}=15.7/14$ against the best-fitting theory. Therefore, we perform the analysis without modeling the RSD effect.

The results are displayed in Figure \ref{fig:errorbar_wtheta}. The subscript `4' and `12' denote the cut-off scales $4 \hMpc$ and $12 \hMpc $, respectively. The red error bars are the main results of this paper shown in Section \ref{sec:result-rcc}.  
For the case of $\gamma_{\rm t}(\theta) + w_4(\theta)$, we obtain $b_{\rm g} = 2.00 \pm 0.14$ and $r_{\rm cc} = 1.06 \pm 0.13$ for the scale cut of $4 \hMpc$, and $b_{\rm g} = 2.03^{+0.17}_{-0.16}$ and $r_{\rm cc} = 1.01^{+0.16}_{-0.15}$ for the scale cut of $12 \hMpc$. 
These results show that $w(\theta)$ of DMASS favors slightly higher galaxy bias than that of CMASS.
We find that the constraining power of the DMASS angular clustering is comparable to the one from the BOSS measurement despite the fairly small survey area of DES Y1 compared to BOSS. This is mainly because the DMASS angular clustering contains smaller scales down to $4 \hMpc$ while the BOSS measurements were obtained over the scales of $r > 40 \hMpc$ \citep{Chuang2017}.    

Next, we constrain parameters by combining all three probes. 
The measurement of BOSS CMASS and the angular clustering of DMASS share the same galaxy clustering bias $b_{\rm g}$ and the tangential shear constrains the lensing galaxy bias $b_{\gamma}$ separately.  We obtain $b_{\rm g} = 2.00 \pm 0.10$ and $r_{\rm cc} = 1.06 \pm 0.10$ for the scale cut of $4 \hMpc$, and $b_{\rm g} = 2.02 \pm 0.11$ and $r_{\rm cc} = 1.03 \pm 0.11$ for the scale cut of $12 \hMpc$. 
Adding angular galaxy clustering improves the constraint on $r_{\rm cc}$  by $23 \%$ for the scale cut of $4 \hMpc$, and $29\%$ for $12 \hMpc$. The improvements on $b_{\rm g}$ are $29 \%$ and $33 \%$ for the scale cut of $4 \hMpc$ and $12 \hMpc$, respectively.

{
\renewcommand{\arraystretch}{1.5}
\begin{table}
\centering
\caption{ 
The constraints of galaxy bias and the cross-correlation parameters with $1\sigma$ errors obtained from the various combinations of data sets. 
 The subscript `4' and `12' denote the cut-off scales $4 \hMpc$ and $12 \hMpc $, respectively. 
}
\label{tab:bias}
\resizebox{0.47\textwidth}{!}{%
\begin{tabular}{@{}llllll@{}}
\toprule
 &                                                  & \multicolumn{1}{c}{$b_{\rm g}$} & \multicolumn{1}{c}{$b_{\gamma}$} & \multicolumn{1}{c}{$r_{\rm cc}$} &  \\ \midrule
 & $\gamma_{\rm t,4}(\theta)$ + BOSS                    & $1.92^{+0.16}_{-0.15}$    & $2.10^{+0.13}_{-0.12}$           & $1.09^{+0.12}_{-0.11}$           &  \\
 & $\gamma_{\rm t,4}(\theta)$ + $w_{4}(\theta)$         & $2.00 \pm 0.14$            & $2.12 \pm 0.14$                  & $1.06 \pm 0.13$                  &  \\
 & $\gamma_{\rm t,4}(\theta)$ + $w_{4}(\theta)$ + BOSS  & $2.00 \pm 0.10$            & $2.13^{+0.12}_{-0.11}$           & $1.06 \pm 0.10$                  &  \\
 & $\gamma_{\rm t,12}(\theta)$ + BOSS                   & $1.92^{+0.16}_{-0.16}$    & $2.04 \pm 0.16$                  & $1.06^{+0.13}_{-0.12}$           &  \\
 & $\gamma_{\rm t,12}(\theta)$ + $w_{4}(\theta)$        & $2.03^{+0.17}_{-0.16}$    & $2.06 \pm 0.17$                  & $1.01^{+0.16}_{-0.15}$            &  \\
 & $\gamma_{\rm t,12}(\theta)$ + $w_{4}(\theta)$ + BOSS & $2.02 \pm 0.11$           & $2.08 \pm 0.13$                  & $1.03 \pm 0.11$                  &  \\ \bottomrule
\end{tabular}%
}
\end{table}
}

%% file: conclusion_ggl.tex
\section{Conclusion}
\label{sec:conclusion}

In this paper, we measured the galaxy-galaxy lensing signal using DMASS lenses and \metacal\  sources. 
To ensure the measured signal is free from various systematic effects, we performed tests for 
the mean cross-component of the shear and the impact of observing conditions. 
We also computed the boost factor and corrected the measured signals for this effect. 
In the scales of $4 \hMpc$ and $12 \hMpc$, we did not find any significant impact of systematics. 
The calibrated signals of tangential shear yield the signal-to-noise ratio of $16.4$ for the scale cut of $4 \hMpc$, and $25.6$ for the scale cut of $12 \hMpc$.

By combining the galaxy-galaxy lensing signals with the BOSS CMASS galaxy clustering measurements, we derived the the cross-correlation coefficient $r_{\rm cc}$ and assessed the equivalence of DMASS and BOSS CMASS. 
We obtained $r_{\rm cc} = 1.09^{+0.12}_{-0.11}$ for the scale cut of $4 \hMpc$ and $r_{\rm cc} = 1.06^{+0.13}_{-0.12}$ for $12 \hMpc$, both are consistent with the ideal value of $r_{\rm cc} = 1$ within $1\sigma$. Adding the angular galaxy clustering of DMASS, the resulting values are $r_{\rm cc}=1.06\pm 0.10$ for the scale cut of $4 \hMpc$ and $r_{\rm cc}=1.03\pm 0.11$ for  $12 \hMpc$. We find that these values agree with the results from other works that utilize the BOSS CMASS galaxies as lenses. Our result indicates that the tangential shear measurement in this work is statistically consistent with the one that would have been measured if BOSS CMASS populates in the DES region. The measured signals will be utilized as the data vector for the joint analysis of DES and BOSS in a forthcoming paper.  

%% file: ack.tex
\section*{Acknowledgements}

AC acknowledges support from NASA grant 15-WFIRST15-0008. During the preparation of this paper, C.H.\ was supported by the Simons Foundation, NASA, and the US Department of Energy.

The figures in this work are produced with plotting routines from matplotlib \citep{matplotlib} and ChainConsumer \citep{ChainConsumer}.
Some of the results in this paper have been derived using the healpy and HEALPix package \citep{HEALPix,healpy}.

Funding for the DES Projects has been provided by the U.S. Department of Energy, the U.S. National Science Foundation, the Ministry of Science and Education of Spain, 
the Science and Technology Facilities Council of the United Kingdom, the Higher Education Funding Council for England, the National Center for Supercomputing 
Applications at the University of Illinois at Urbana-Champaign, the Kavli Institute of Cosmological Physics at the University of Chicago, 
the Center for Cosmology and Astro-Particle Physics at the Ohio State University,
the Mitchell Institute for Fundamental Physics and Astronomy at Texas A\&M University, Financiadora de Estudos e Projetos, 
Funda{\c c}{\~a}o Carlos Chagas Filho de Amparo {\`a} Pesquisa do Estado do Rio de Janeiro, Conselho Nacional de Desenvolvimento Cient{\'i}fico e Tecnol{\'o}gico and 
the Minist{\'e}rio da Ci{\^e}ncia, Tecnologia e Inova{\c c}{\~a}o, the Deutsche Forschungsgemeinschaft and the Collaborating Institutions in the Dark Energy Survey. 

The Collaborating Institutions are Argonne National Laboratory, the University of California at Santa Cruz, the University of Cambridge, Centro de Investigaciones Energ{\'e}ticas, 
Medioambientales y Tecnol{\'o}gicas-Madrid, the University of Chicago, University College London, the DES-Brazil Consortium, the University of Edinburgh, 
the Eidgen{\"o}ssische Technische Hochschule (ETH) Z{\"u}rich, 
Fermi National Accelerator Laboratory, the University of Illinois at Urbana-Champaign, the Institut de Ci{\`e}ncies de l'Espai (IEEC/CSIC), 
the Institut de F{\'i}sica d'Altes Energies, Lawrence Berkeley National Laboratory, the Ludwig-Maximilians Universit{\"a}t M{\"u}nchen and the associated Excellence Cluster Universe, 
the University of Michigan, NFS's NOIRLab, the University of Nottingham, The Ohio State University, the University of Pennsylvania, the University of Portsmouth, 
SLAC National Accelerator Laboratory, Stanford University, the University of Sussex, Texas A\&M University, and the OzDES Membership Consortium.

Based in part on observations at Cerro Tololo Inter-American Observatory at NSF's NOIRLab (NOIRLab Prop. ID 2012B-0001; PI: J. Frieman), which is managed by the Association of Universities for Research in Astronomy (AURA) under a cooperative agreement with the National Science Foundation.

The DES data management system is supported by the National Science Foundation under Grant Numbers AST-1138766 and AST-1536171.
The DES participants from Spanish institutions are partially supported by MICINN under grants ESP2017-89838, PGC2018-094773, PGC2018-102021, SEV-2016-0588, SEV-2016-0597, and MDM-2015-0509, some of which include ERDF funds from the European Union. IFAE is partially funded by the CERCA program of the Generalitat de Catalunya.
Research leading to these results has received funding from the European Research
Council under the European Union's Seventh Framework Program (FP7/2007-2013) including ERC grant agreements 240672, 291329, and 306478.
We  acknowledge support from the Brazilian Instituto Nacional de Ci\^encia
e Tecnologia (INCT) do e-Universo (CNPq grant 465376/2014-2).

This manuscript has been authored by Fermi Research Alliance, LLC under Contract No. DE-AC02-07CH11359 with the U.S. Department of Energy, Office of Science, Office of High Energy Physics.

Funding for SDSS-III has been provided by the Alfred P. Sloan Foundation, the Participating Institutions, the National Science Foundation, and the U.S. Department of Energy Office of Science. The SDSS-III web site is \url{http://www.sdss3.org/}.

SDSS-III is managed by the Astrophysical Research Consortium for the Participating Institutions of the SDSS-III Collaboration including the University of Arizona, the Brazilian Participation Group, Brookhaven National Laboratory, Carnegie Mellon University, University of Florida, the French Participation Group, the German Participation Group, Harvard University, the Instituto de Astrofisica de Canarias, the Michigan State/Notre Dame/JINA Participation Group, Johns Hopkins University, Lawrence Berkeley National Laboratory, Max Planck Institute for Astrophysics, Max Planck Institute for Extraterrestrial Physics, New Mexico State University, New York University, Ohio State University, Pennsylvania State University, University of Portsmouth, Princeton University, the Spanish Participation Group, University of Tokyo, University of Utah, Vanderbilt University, University of Virginia, University of Washington, and Yale University.

This work used resources at the Owens Cluster at the Ohio Supercomputer Center  \citep{OSC} and the Duke Compute Cluster (DCC) at Duke University.